\documentclass[10pt,aps,prd,onecolumn,superscriptaddress,nofootinbib,longbibliography,floatfix]{revtex4-2}
\usepackage{amsmath,amssymb,amsfonts,bm,mathtools}
\usepackage{graphicx,booktabs,array}
\usepackage{placeins}
\usepackage{microtype}
\usepackage{xcolor}
\usepackage{hyperref}
\usepackage{slashed}
\usepackage{orcidlink}
\usepackage{subcaption}
\hypersetup{pdftitle={Gravitational Form Factors and Transverse Mechanical Structure of the Charged Kaon},pdfauthor={Parashmani Thakuria, Madhurjya Lalung, Jayanta Kumar Sarma},colorlinks=true,linkcolor=blue!45!black,citecolor=blue!45!black,urlcolor=blue!45!black,hypertexnames=false}
\graphicspath{{./}{figures/}}

\newcommand{\paperfigure}[2]{%
  \includegraphics[width=#1]{#2}%
}

\allowdisplaybreaks

\begin{document}
\raggedbottom

\title{Gravitational Form Factors and Transverse Mechanical Structure of the Charged Kaon in a Two-Fock-Sector Light-Front Hamiltonian}

\author{Parashmani Thakuria\orcidlink{0000-0002-5652-7835}}
\email{parasht@tezu.ernet.in}
\affiliation{Department of Physics, School of Sciences, Tezpur University, Tezpur, India, Pin-784028}

\author{Madhurjya Lalung\orcidlink{0000-0002-7763-5050}}
\email{mlalung2016@gmail.com}
\affiliation{Department of Physics, Nagaon University, Nagaon, India}

\author{Jayanta Kumar Sarma}
\email{jks@tezu.ernet.in}
\affiliation{Department of Physics, School of Sciences, Tezpur University,\ Tezpur,\ India,\ Pin-784028\ }

\date{\today}

\begin{abstract}
We study the electromagnetic and gravitational form factors of the charged kaon in a light-front Hamiltonian model with coupled $|u\bar s\rangle$ and $|u\bar s g\rangle$ Fock sectors. We solve the coupled Hamiltonian in a longitudinal Jacobi and transverse harmonic-oscillator basis, with the interaction strength fixed by the kaon mass. The momentum form factors are extracted from $T^{++}$, while the stress form factors are obtained from the transverse shear projection of the QCD Belinfante energy-momentum tensor, including diagonal, sector-changing, and constrained-field contributions. The strange antiquark carries the largest fraction of the kaon's longitudinal momentum, whereas the gluon carries only $2.05\%$. Despite this small momentum fraction, the gluon contributes $42.4\%$ of $|D_K(0)|$, for which we obtain $D_K(0)=-2.03610$. The charge and transverse momentum radii are $0.5178\pm0.0278\,\mathrm{fm}$ and $0.3394\pm0.0113\,\mathrm{fm}$, respectively, with uncertainties estimated from mass and basis variations. We obtain the transverse mechanical distributions using two ultraviolet continuations of the directly calculated $D$ form factors beyond $Q^2=2\,\mathrm{GeV}^2$. Both continuations give positive pressure at small transverse distances and negative pressure at larger distances. The corresponding transverse mechanical radii are $0.4616\pm0.0130\,\mathrm{fm}$ and $0.4768\pm0.0132\,\mathrm{fm}$, both larger than the transverse momentum radius. The central pressure changes substantially between the two continuations, whereas the pressure zero and mechanical radii vary much less.
\end{abstract}

\maketitle

\section{Introduction}
\label{sec:introduction}

Gravitational form factors (GFFs) parameterize hadron matrix
elements of the energy-momentum tensor (EMT) and describe
how quarks and gluons carry momentum and contribute to
internal stresses \cite{Pagels1966EMT,Teryaev2016GFF}.
Their relation to Mellin moments of generalized parton
distributions (GPDs) connects these properties to hard
exclusive reactions
\cite{Ji1997GPD,Radyushkin1997DVCS,Diehl2003Review,
BelitskyRadyushkin2005}.
For a spin-zero hadron, $A(t)$ determines the
longitudinal-momentum distribution, while $D(t)$ encodes
the shear part of the stress tensor
\cite{Polyakov2003Forces,PolyakovSchweitzer2018,
Lorce2019Mechanics}.
The impact-parameter interpretation of GPDs at zero
skewness \cite{Burkardt2000Impact,Burkardt2003Impact}
and the light-front formulation of stress densities
\cite{FreeseMiller2021} provide a framework for studying
transverse momentum distributions, pressure, shear,
and mechanical radii.

Experimental constraints have mainly come from proton
deeply virtual Compton scattering, which provides
information on the quark $D$ term and associated forces
\cite{Burkert2018Proton,Burkert2021Shear,
BurkertEtAl2023Review}.
Near-threshold $J/\psi$ photoproduction has also been
used to extract gluonic GFFs within reaction models
\cite{Duran2023Gluon}.
The transverse interpretation of the extracted pressure
has been studied in Ref.~\cite{FreeseMiller2021Empirical}.
For mesons, Belle measurements of
$\gamma^*\gamma\to\pi^0\pi^0$ have enabled extractions
of pion generalized distribution amplitudes and
timelike GFFs \cite{Masuda2016Belle,Kumano2018TwoPhoton}.
Direct constraints on kaon GFFs remain limited.

The pion and kaon connect hadron structure to dynamical
chiral symmetry breaking and explicit quark-mass effects
\cite{Roberts2021Mesons,Donoghue1991Pion}.
The kaon's light-strange mass asymmetry makes it
particularly useful for studying flavor-dependent
momentum and spatial distributions.
GPD-based studies have examined this asymmetry through
electromagnetic form factors, gravitational radii,
and pressure profiles
\cite{Zhang2021MesonGPD,Raya2022MesonGPD,
SonHutauruk2025GPD}.
Continuum calculations with dressed currents have also
obtained pion and kaon electromagnetic and gravitational
form factors within a common framework
\cite{XuDing2024Kaon}.Lattice calculations provide gluon GFFs for the pion and other hadrons
\cite{ShanahanDetmold2019Gluon,Pefkou2022Hadron},
the quark and gluon decomposition of pion GFFs
\cite{Hackett2023PionLattice}, and momentum fractions
in the pion and kaon \cite{Alexandrou2025PionKaonMom}.
Complementary descriptions use Dyson-Schwinger and
Bethe-Salpeter equations \cite{Mezrag2014PionGPD},
chiral quark models and contact interactions
\cite{Broniowski2008Pion,Sultan2024Pseudoscalar},
dispersion relations \cite{CaoGuo2025Dispersive},
holographic models
\cite{AbidinCarlson2008Axial,LiuWatanabe2025Kaon,
LiVary2024PionStress}, and the QCD instanton vacuum
\cite{LiuShuryak2024I,LiuShuryak2024II}.
These approaches provide benchmarks for the form factors
and mechanical distributions, with comparisons depending
on the retained degrees of freedom, current operators,
and renormalization prescription.

The light-front Hamiltonian approach describes hadrons
through Fock-state wave functions
\cite{BrodskyPauliPinsky1998}, from which parton
distributions and current matrix elements can be
evaluated \cite{Diehl2001Overlap,BrodskyHwang2001}.
Basis Light-Front Quantization (BLFQ) implements this
approach through a finite basis expansion
\cite{Vary2010BLFQ}.
Jacobi and harmonic-oscillator bases have been applied
to heavy quarkonia and light mesons
\cite{LiMarisZhao2016,JiaVary2019NJL,Lan2020Kaon},
and light-meson calculations have been extended to
the coupled $q\bar q\oplus q\bar q g$ space
\cite{Lan2022DynamicalGluon}.
BLFQ studies of GFFs include the proton
\cite{Nair2024ProtonBLFQ} and a recent valence-sector
calculation for the pion and kaon
\cite{Sain2026KaonBLFQ}.
The latter identifies small-$x$ sensitivity in the
transverse-current extraction of $D(t)$.
Related studies examine the choice of EMT component
and the treatment of interactions
\cite{CaoEtAl2024LFEMT,Hu2025ShearStress,
Choi2025PionLFQM}.
An explicit gluon sector allows its momentum and
stress contributions to be resolved alongside those
of the quark and antiquark.

We examine how an explicit gluon component contributes to the momentum and stress distributions of the charged kaon within a common bound-state calculation. We solve the coupled $|u\bar s\rangle\oplus|u\bar s g\rangle$ Hamiltonian in a Jacobi-harmonic-oscillator basis with full transverse support. The Hamiltonian includes valence confinement, an effective gluon kinetic mass, and a smooth invariant-mass factor in the transition vertices. After calibrating the coupling to the kaon mass, we use the resulting eigenstates to calculate the electromagnetic form factor and the separate $u$, $\bar s$, and gluon GFFs. The projected symmetric QCD Belinfante current retains diagonal, sector-changing, and constrained-field terms. We assess mass and basis sensitivity through variations of the eigenstates and examine the mechanical reconstruction using two ultraviolet continuations of the directly calculated $D$ form factors.

Section~\ref{sec:theory} presents the Hamiltonian
and basis expansion.
Section~\ref{sec:currents} defines the form factors
and mechanical observables.
The results and conclusions are given in
Secs.~\ref{sec:finalResults} and \ref{sec:conclusion}.
Appendices~\ref{app:kernels} and \ref{app:transforms}
contain the interaction kernels and analytic
transform formulas.

\section{Light-front Hamiltonian and kaon wave functions}
\label{sec:theory}

We study the charged kaon in a light-front Hamiltonian
framework, retaining the valence $|u\bar s\rangle$ and
one-gluon $|u\bar s g\rangle$ Fock sectors
\cite{BrodskyPauliPinsky1998}. A similar truncation,
including valence confinement, has been used in light-meson calculations~\cite{Lan2022DynamicalGluon}.

\subsection{Two-sector Hamiltonian and light-front wave functions}
\label{sec:theoryHamiltonian}

We work in light-cone gauge and use the conventions
\cite{BrodskyPauliPinsky1998,SrivastavaBrodsky}
\begin{equation}
 x^\pm=\frac{x^0\pm x^3}{\sqrt2},\qquad
 p^2=2p^+p^-\bm p_\perp^2,\qquad A^+=0.
 \label{eq:lfConventionsNew}
\end{equation}
The mass-squared operator is denoted by $H_{\rm LF}$, and its kaon
eigenvalue is $M_K^2$. We retain the valence and one-dynamical-gluon
components,
\begin{equation}
 |\Psi_K\rangle=|\psi_2\rangle+|\psi_3\rangle,
 \quad |\psi_2\rangle=|u\bar s\rangle,\quad
 |\psi_3\rangle=|u\bar s g\rangle.
 \label{eq:fockstateNew}
\end{equation}
The projected eigenproblem is
\begin{equation}
 \begin{pmatrix}
 H_{22}&H_{23}\\
 H_{32}&H_{33}
 \end{pmatrix}
 \begin{pmatrix}\psi_2\\\psi_3\end{pmatrix}
 =
 M_K^2
 \begin{pmatrix}\psi_2\\\psi_3\end{pmatrix},
 \qquad P_2+P_3=1.
 \label{eq:mastereig}
\end{equation}
The two Fock-sector amplitudes are therefore obtained
simultaneously from the coupled eigenvalue problem.

The canonical Hamiltonian is organized as
$H_{\rm QCD}=H_0+gV_1+g^2V_2$, where the interaction terms include
the contributions generated by eliminating the nondynamical fields
\cite{BrodskyPauliPinsky1998,SrivastavaBrodsky}.
Our projected operator is
\begin{equation}
 H_{\rm prod}=
 \begin{pmatrix}
 H_{22}^{\rm QCD}+V_{\rm conf}
 &g\,\mathcal V_{23}^{(\lambda_V,m_g)}\\
 g\,\mathcal V_{32}^{(\lambda_V,m_g)}
 &H_{33}^{\rm QCD}(m_g)
 \end{pmatrix},
 \quad
 \mathcal V_{23}=\mathcal V_{32}^\dagger.
 \label{eq:Hprod}
\end{equation}

We use normalized color-singlet amplitudes
$C_2^{ij}=\delta_{ij}/\sqrt3$ and
$C_3^{ija}=(t^a)_{ij}/2$, with
$\mathrm{tr}(t^at^b)=\delta^{ab}/2$.
The generator normalization and quadratic Casimirs follow the
standard $SU(3)$ identities \cite{Haber2021Color}.
In the three-body singlet, the $u\bar s$ pair is a color octet.
Using
$2\bm T_i\!\cdot\bm T_j=C_{ij}-C_i-C_j$,
where $C_{ij}$ is the quadratic Casimir of the combined pair,
together with $C_F=4/3$ and $C_A=3$, gives
\begin{equation}
 \bm T_u\!\cdot\bm T_{\bar s}=
 \begin{cases}
 -4/3,&u\bar s,\\
 1/6,&u\bar s g,
 \end{cases}
 \qquad
 \bm T_{u,\bar s}\!\cdot\bm T_g=-\frac32.
 \label{eq:paircolors}
\end{equation}
These products determine the relative color strengths of the
instantaneous interactions. In our normal-ordering and phase
conventions, the emission vertices have the color factors
\begin{equation}
 {\cal C}_{u\to ug}=-\frac{2}{\sqrt3},\qquad
 {\cal C}_{\bar s\to\bar s g}=\frac{2}{\sqrt3}.
 \label{eq:cubiccolors}
\end{equation}
The relative sign follows from the fundamental and
antifundamental color generators.

At $\bm P_\perp=0$, the two-body intrinsic variables satisfy
$x_u=x$, $x_{\bar s}=1-x$, and
$\bm k_{\bar s\perp}=-\bm k_{u\perp}=-\bm k_\perp$.
The pseudoscalar two-parton state contains helicity-antialigned
and helicity-aligned amplitudes with $L_z=0$ and $|L_z|=1$,
respectively \cite{JiMaYuan2004}.
We write these amplitudes in terms of $U(x,k)$ and $V(x,k)$ as
\begin{equation}
 \begin{aligned}
 \psi_{2,+-}&=U/\sqrt2,
 &\psi_{2,-+}&=-U/\sqrt2,\\
 \psi_{2,++}&=e^{-i\phi_k}V/\sqrt2,
 &\psi_{2,-}&=e^{i\phi_k}V/\sqrt2.
 \end{aligned}
 \label{eq:psi2explicit}
\end{equation}
Thus $U$ and $V$ carry orbital projections $|m|=0$ and $|m|=1$.
In the wave-function normalization used here, the valence
probability is
\begin{equation}
 P_2=2\pi\int_0^1\!dx\int_0^\infty\!k\,dk\;
 \bigl(|U(x,k)|^2+|V(x,k)|^2\bigr).
 \label{eq:P2}
\end{equation}

For the three-body sector,
$x_1+x_2+x_3=1$ and
$\bm k_{1\perp}+\bm k_{2\perp}+\bm k_{3\perp}=0$.
The labels $1,2,3$ refer to the $u$ quark, strange antiquark,
and transverse gluon. For each helicity configuration
$\alpha=(h_1,h_2,\sigma)$, with $h_i=\pm\tfrac12$ and
$\sigma=\pm1$, the condition $J^z=0$ gives
\begin{equation}
 \begin{aligned}
 m_\alpha&=-(h_1+h_2+\sigma),\\
 \psi_{3,\alpha}
 &=e^{im_\alpha\phi_1}
 f_\alpha(x_1,x_2,k_1,k_2,\phi_2-\phi_1).
 \end{aligned}
 \label{eq:phase3}
\end{equation}
This separation of helicity and orbital angular momentum follows
the light-front Fock-state construction
\cite{BrodskyHwang2001,JiMaYuan2004}.
We retain all eight combinations of $(h_1,h_2,\sigma)$,
with $\sigma=\pm1$ denoting the two physical transverse
gluon polarizations. The parameter $m_g$ introduced
below enters the kinetic energy and does not introduce a
longitudinal gluon polarization.

The one-gluon probability is
\begin{equation}
 \begin{aligned}
 P_3&=\sum_\alpha
 \int_{x_1+x_2<1}\!dx_1dx_2
 \int\!d^2k_1d^2k_2\,|\psi_{3,\alpha}|^2,\\
 P_2+P_3&=1.
 \end{aligned}
 \label{eq:P3}
\end{equation}
Equations~\eqref{eq:P2} and \eqref{eq:P3} define the normalization
used in the basis projection and current overlaps.

The free invariant masses follow from the light-front
kinetic-energy~\cite{BrodskyPauliPinsky1998} expression
$\mathcal M_N^2=\sum_i(m_i^2+k_{i\perp}^2)/x_i$,

\begin{align}
 \mathcal M_2^2
 &=\frac{m_u^2+k_\perp^2}{x}
 +\frac{m_s^2+k_\perp^2}{1-x},
 \label{eq:D2}\\
 \mathcal M_3^2
 &=\frac{m_u^2+k_{1\perp}^2}{x_1}
 +\frac{m_s^2+k_{2\perp}^2}{x_2}
 +\frac{m_g^2+k_{3\perp}^2}{x_3}.
 \label{eq:D3}
\end{align}
The positive $m_g^2/x_3$ term suppresses the gluon endpoint
within the retained sector. A phenomenological gluon kinetic
mass is also used in the light-meson Hamiltonian of
Ref.~\cite{Lan2022DynamicalGluon}.
Here $m_g$ is a parameter of the finite-resolution Hamiltonian.
We retain the canonical $q\to qg$ vertex numerator.

The valence confinement combines transverse holographic
confinement with a longitudinal differential operator
\cite{Brodsky2015Holography,LiMarisZhao2016,
LiMarisVary2017,Lan2022DynamicalGluon}:
\begin{equation}
 \begin{aligned}
 V_{\rm conf}
 &=\kappa^4\biggl[
 x(1-x)\bm r_\perp^2-\frac{1}{(m_u+m_s)^2}
 \partial_x\bigl(x(1-x)\partial_x\bigr)
 \biggr],
 \bm r_\perp^2\longleftrightarrow-\nabla_{\bm k}^2.
 \end{aligned}
 \label{eq:Vconf}
\end{equation}
It acts only in $H_{22}$.

For the sector-changing blocks, we introduce a smooth
invariant-mass factor on both emission routes,
\begin{equation}
 \mathcal V_{32}
 =F_{\lambda_V}(\Delta\mathcal M^2)
 \mathcal V_{32}^{\rm can},
 \qquad
 F_{\lambda_V}(z)
 =\exp\left[-\left(\frac{z}{\lambda_V^2}\right)^2\right].
 \label{eq:vertexReg}
\end{equation}
Gaussian suppression of invariant-mass changes also appears
in the renormalization group procedure for effective particles
\cite{Glazek2012Effective}.
Here we use this functional form to define the vertex resolution
of our finite Hamiltonian. The associated renormalization-group
flow and counterterm construction are beyond the present
calculation.

For emission from the $u$ and $\bar s$ lines, the invariant-mass
differences are
\begin{equation}
 \begin{split}
 \Delta\mathcal M_u^2
 &=\mathcal M_3^2(m_g)
 -\mathcal M_2^2(1-x_2,|\bm k_2|),\\
 \Delta\mathcal M_{\bar s}^2
 &=\mathcal M_3^2(m_g)
 -\mathcal M_2^2(x_1,|\bm k_1|).
 \end{split}
 \label{eq:deltaMroutes}
\end{equation}
The same $m_g$ enters the three-body kinetic energy and these
mass differences. The resolution factor acts in the
Hamiltonian and is not applied to the EMT insertion.

Within the retained Fock space, the direct interactions contain
quark-antiquark, quark-gluon, and antiquark-gluon pair kernels.
We define the regulated constrained-field denominators by
\begin{equation}
 R_{1,\epsilon}(z)=\frac{z}{z^2+\epsilon^2},
 \qquad
 R_{2,\epsilon}(z)=\frac{z^2}{(z^2+\epsilon^2)^2}.
 \label{eq:R12}
\end{equation}
The gluonic and mixed contributions generated by eliminating
the nondynamical fields are retained according to the canonical
light-cone-gauge construction \cite{SrivastavaBrodsky}.
The finite-$\epsilon$ prescription defined in
Eq.~\eqref{eq:R12} is used throughout the kernel integrals.

The emission kernels, three-body pair interactions, and
projected block actions are given in
Appendix~\ref{app:kernels}. They use the same regulator and
satisfy $\mathcal V_{23}=\mathcal V_{32}^{\dagger}$.

\subsection{Analytic basis and eigenstates}
\label{sec:theoryBasis}

We solve the eigenproblem by Rayleigh-Ritz projection onto
longitudinal Jacobi functions and transverse
harmonic-oscillator (HO) functions, following the
basis-Hamiltonian strategy of BLFQ \cite{Vary2010BLFQ}.
Related expansions have been used for heavy quarkonia and
light mesons
\cite{LiMarisZhao2016,LiMarisVary2017,
JiaVary2019NJL,Lan2020Kaon}.
Our expansion includes both the valence and one-gluon sectors.
At fixed truncation, the Hamiltonian is a finite Hermitian
operator in coefficient space. Quadrature is used to evaluate
its matrix elements. The transverse HO basis spans the full plane
$\mathbb R^2$.

The valence longitudinal functions diagonalize the quark-mass
terms and the longitudinal part of $V_{\rm conf}$. Defining
\begin{equation}
 p=\frac{m_u(m_u+m_s)}{\kappa^2},
 \qquad
 r=\frac{m_s(m_u+m_s)}{\kappa^2},
 \label{eq:prdef}
\end{equation}
the Jacobi construction
\cite{LiMarisZhao2016,LiMarisVary2017} gives
\begin{align}
 \chi_\ell(x)
 &=\mathcal N_\ell(2r,2p)
 x^p(1-x)^rP_\ell^{(2r,2p)}(2x-1),
 \label{eq:chiJacobi}\\
 \Biggl[
 \frac{m_u^2}{x}+\frac{m_s^2}{1-x}
 -\frac{\kappa^4}{(m_u+m_s)^2}
 \partial_x\bigl(x(1-x)\partial_x\bigr)
 \Biggr]\chi_\ell(x)
 &=\frac{\kappa^4}{(m_u+m_s)^2}
 (\ell+p+r)\times
 (\ell+p+r+1)\chi_\ell(x).
 \label{eq:LongExactEig}
\end{align}
The constants $\mathcal N_\ell$ enforce
$\int_0^1dx\,\chi_\ell\chi_{\ell'}=\delta_{\ell\ell'}$.

At transverse oscillator scale $\beta$, we use the normalized
full-plane HO functions
\cite{Vary2010BLFQ,LiMarisZhao2016}
\begin{equation}
 \begin{aligned}
 \phi_{nm}(\bm k;\beta)
 &=\frac1\beta
 \sqrt{\frac{n!}{\pi\Gamma(n+|m|+1)}}
 \left(\frac{k}{\beta}\right)^{|m|}
 e^{-k^2/(2\beta^2)}
 L_n^{|m|}(k^2/\beta^2)e^{im\phi_k},
 \end{aligned}
 \label{eq:rawHO}
\end{equation}
which satisfy
$\int_{\mathbb R^2}d^2k\,
\phi_{nm}^*\phi_{n'm'}=\delta_{nn'}\delta_{mm'}$.
Their Gaussian factor supplies the large-momentum falloff
without imposing a hard transverse boundary.

The valence expansion is
\begin{equation}
 \begin{aligned}
 \psi_{2,h\bar h}(x,\bm k)
 &=\sum_{\substack{n,\ell,\eta\\
 2n+s_\eta+\ell\le N_{\max}^{(2)}}}
 c_{n\ell\eta}^{(2)}\,\chi_\ell(x)\times R_{n,s_\eta}(k;\beta_2)
 \Xi_{\eta,h\bar h}(\phi_k),
 \qquad \eta=U,V.
 \end{aligned}
 \label{eq:psi2BasisExpansion}
\end{equation}
Here $s_U=0$, $s_V=1$, and $\Xi_\eta$ contains the spin
combination in Eq.~\eqref{eq:psi2explicit}.
We choose
$\beta_2=\kappa\sqrt{x_0(1-x_0)}$ with
$x_0=m_u/(m_u+m_s)$.

In the three-body sector, we map the longitudinal simplex to
$0<u,v<1$ through
\begin{equation}
 \begin{aligned}
 x_1&=u,\qquad x_2=(1-u)v,\\
 x_3&=(1-u)(1-v),\\
 dx_1dx_2&=(1-u)\,du\,dv.
 \end{aligned}
 \label{eq:uvmap}
\end{equation}
We construct the product-Jacobi basis
$X_{\ell j}(u,v)=U_\ell(u)V_j(v)$, where
\begin{align}
 U_\ell(u)
 &=\mathcal N_\ell(2r+2c_3+1,2p)\times
 u^p(1-u)^{r+c_3}
 P_\ell^{(2r+2c_3+1,2p)}(2u-1),\\
 V_j(v)
 &=\mathcal N_j(2c_3,2r)
 v^r(1-v)^{c_3}\times
 P_j^{(2c_3,2r)}(2v-1).
 \label{eq:threeBodyJacobiNew}
\end{align}
The construction uses one-dimensional Jacobi orthogonality
\cite{DLMFOrthogonalPolynomials}, with the Jacobian of
Eq.~\eqref{eq:uvmap} included in the $u$ weight.
The product functions are orthonormal under
$du\,dv\,(1-u)$.
The parameter $c_3>0$ controls the longitudinal basis shape.

We use two full-plane HO factors with
$\beta_3=s_\beta\beta_2$ and impose
\begin{equation}
 m_1+m_2=m_\alpha,\qquad
 2n_1+|m_1|+2n_2+|m_2|+\ell+j
 \le N_{\max}^{(3)}.
 \label{eq:Nmax3}
\end{equation}
The expansion coefficients and the truncations
$(N_{\max}^{(2)},N_{\max}^{(3)})$ specify the wave functions.
Orthogonality gives
\begin{equation}
 \begin{aligned}
 P_2&=\sum_i|c_i^{(2)}|^2,\quad
 P_3=\sum_A|c_A^{(3)}|^2, \quad
 \|c_2\|^2+\|c_3\|^2=1.
 \end{aligned}
 \label{eq:coeffNorm}
\end{equation}

We factorize the $H_{33}$ contractions by helicity and active
pair, allowing the Hamiltonian to act on a coefficient vector
without storing a dense matrix.
The production basis contains $N_2=91$ and $N_3=45018$
coefficients, giving a total dimension of $45109$.
A restarted matrix-free eigensolver determines the lowest
pseudoscalar state. We monitor the normalized residual
\begin{equation}
 \mathcal R_{\rm eig}
 =\frac{\|Hc-M_K^2c\|_2}
 {\max(1,|M_K^2|)}.
 \label{eq:lanczosstop}
\end{equation}
This construction makes the coupled-sector calculation
practical at the stated basis size.

\section{Electromagnetic and gravitational form factors}
\label{sec:currents}

All matrix elements are evaluated between the eigenstates defined in Sec.~\ref{sec:theory}. For the EMT we adopt the projected symmetric QCD Belinfante current. Confinement and the vertex-resolution factor enter the Hamiltonian; the adopted prescription does not include their additional metric-response currents \cite{CaoEtAl2024LFEMT,LiWangVary2024EMT,CaoLiVary2023Scalar}.

\subsection{Electromagnetic current}
\label{sec:theoryEM}

The electromagnetic current provides a comparison with charged-kaon scattering data using the same bound-state wave functions. Let $J_u$ and $J_{\bar s}$ denote the unit-charge constituent overlaps, normalized by $J_u(0)=J_{\bar s}(0)=1$. The charged-kaon form factor is
\begin{equation}
 F_{K^+}(-Q^2)=\frac23J_u(-Q^2)+\frac13J_{\bar s}(-Q^2),
 \qquad F_{K^+}(0)=1.
 \label{eq:EMformfactor}
\end{equation}
The electromagnetic overlap uses the recoil shifts in
Eq.~\eqref{eq:diagShiftGeneral}, with the constituent charge
replacing the parton momentum fraction. Both Fock sectors contribute through the charged fermions. The gluon affects this observable through the three-body wave function and its correlations with the quark and antiquark.

The electromagnetic charge radius is defined by
\begin{equation}
 r_{\rm ch}^2=-6(\hbar c)^2
 \left.\frac{dF_{K^+}(-Q^2)}{dQ^2}\right|_{Q^2=0},
 \label{eq:EMradius}
\end{equation}

\subsection{Spin-zero energy-momentum tensor and momentum form factors}
\label{sec:theoryA}

We define the symmetric QCD Belinfante tensor, with $D_\mu=\partial_\mu-igA_\mu$, as in the standard gauge-invariant EMT analysis \cite{HattaRajanTanaka2018,ShanahanDetmold2019Gluon}:
\begin{align}
 T_q^{\mu\nu}&=\frac{i}{4}\sum_{f=u,s}\bar q_f
 (\gamma^\mu\overleftrightarrow D^\nu+\gamma^\nu\overleftrightarrow D^\mu)q_f,\\
 T_g^{\mu\nu}&=-F_a^{\mu\alpha}F^{\nu}{}_{\alpha a}
 +\frac14g^{\mu\nu}F_a^{\alpha\beta}F_{\alpha\beta a}.
 \label{eq:belinfante}
\end{align}
The quark, antiquark and gluon operators are evaluated in the
same two-sector Hilbert space, but their individual matrix elements need not be conserved. Moreover, the separation between quark and gluon contributions depends on the renormalization prescription\cite{HattaRajanTanaka2018,PolyakovSon2018,ShanahanDetmold2019Gluon}. For a spin-zero state of mass $M_K$, with $\bar P=(P'+P)/2$, $\Delta=P'-P$ and $t=\Delta^2$, the decomposition can be written as
\begin{align}
 \langle P'|T_a^{\mu\nu}|P\rangle
 ={}&2\bar P^\mu\bar P^\nu A_a(t)
 +\frac12(\Delta^\mu\Delta^\nu-g^{\mu\nu}\Delta^2)D_a(t)
+2M_K^2g^{\mu\nu}\bar C_a(t)
 +\mathcal M_{a,\mathrm{spur}}^{\mu\nu}.
 \label{eq:spinzeroGFF}
\end{align}
The last term accounts for additional structures that can
arise from the light-front truncation. The separate $\bar C_a$ need not vanish. The shear projection below removes the isotropic $\bar C_a$ term without imposing separate conservation on each partonic current \cite{CaoEtAl2024LFEMT,Hu2025ShearStress}.

We choose the symmetric Drell-Yan frame $\Delta^+=\Delta^-=0$, $\bar{\bm P}_\perp=0$, with $\bm\Delta_\perp=\bm q_\perp$ and $t=-Q^2$. The plus-plus component isolates the partonic momentum form factor:
\begin{equation}
 A_a(-Q^2)=\frac{\langle P'|T_a^{++}|P\rangle}{2(P^+)^2},
 \qquad A_K=\sum_{a=u,\bar s,g}A_a.
 \label{eq:AfromTpp}
\end{equation}
The $T^{++}$ matrix element is diagonal in Fock space and can
be evaluated as a light-front wave-function overlap
\cite{BrodskyDeTeramond2008GFF,BrodskyHwang2001}. For $N=2,3$ the overlap has a common form
\begin{equation}
 \begin{aligned}
 A_a^{(N)}(-Q^2)&=\sum_{\beta_N}\int d\Gamma_N\,x_a\times\psi_{N,\beta_N}^*(\{x_i,\bm k_i'\})
 \psi_{N,\beta_N}(\{x_i,\bm k_i\}).
 \end{aligned}
 \label{eq:AoverlapGeneral}
\end{equation}
where $d\Gamma_N$ denotes the normalized intrinsic LFWF measure in that sector. The quark and antiquark receive contributions from $N=2,3$, while $A_g=A_g^{(3)}$. The intrinsic transverse shifts follow from momentum conservation exactly:
\begin{equation}
 \bm k_{a\perp}'=\bm k_{a\perp}+(1-x_a)\bm q_\perp,
 \qquad \bm k_{j\perp}'=\bm k_{j\perp}-x_j\bm q_\perp\ (j\ne a).
 \label{eq:diagShiftGeneral}
\end{equation}
They preserve $\sum_i\bm k_{i\perp}'=0$ and require neither an approximate recoil prescription nor extrapolation beyond the stored basis. For the three-body coordinates of Eq.~\eqref{eq:uvmap}, the shifts of the two independent transverse momenta are
\begin{equation}
\begingroup
\renewcommand{\arraystretch}{1.35}
\setlength{\arraycolsep}{11pt}
\begin{array}{c|cc}
\hline
\text{Struck parton}
& \bm{k}_{1\perp}^{\prime}
& \bm{k}_{2\perp}^{\prime}
\\
\hline
u
& \bm{k}_1+(1-x_1)\bm{q}
& \bm{k}_2-x_2\bm{q}
\\
\bar{s}
& \bm{k}_1-x_1\bm{q}
& \bm{k}_2+(1-x_2)\bm{q}
\\
g
& \bm{k}_1-x_1\bm{q}
& \bm{k}_2-x_2\bm{q}
\\
\hline
\end{array}
\endgroup
\label{eq:threeBodyAshifts}
\end{equation}
We evaluate the shifted wave functions through analytic HO translation matrix elements. For a transverse shift $\bm\delta$, a representative matrix element is
\begin{equation}
 \mathcal D_{nm,n'm'}(\bm\delta;\beta)=\int_{\mathbb R^2}\!d^2k\;
 \phi_{nm}^*(\bm k+\bm\delta;\beta)
 \phi_{n'm'}(\bm k;\beta),
 \label{eq:HOshiftDef}
\end{equation}
which has a closed Laguerre-polynomial representation. The full-plane transverse basis allows these shifted overlaps to be evaluated at arbitrary momentum transfer. All transfer directions $\phi_q$ are included when evaluating the $D$-term shear matrix elements. At the origin, the good-current sum rule becomes
\begin{equation}
 A_K(0)=\langle x\rangle_u+\langle x\rangle_{\bar s}
 +\langle x\rangle_g=P_2+P_3=1.
 \label{eq:AsumRule}
\end{equation}
The forward fractions can be confronted with lattice calculations only after addressing the effective model scale and quark-gluon operator mixing \cite{Yang2018PionGluon,ShanahanDetmold2019Gluon,Alexandrou2025PionKaonMom}. Any small deviation from the sum rule in the numerical $T^{++}$ quadrature is treated as a closure error. To compare the form-factor curves across eigenstates, we normalize each state separately, $\widehat A_{a,s}=A_{a,s}^{\rm raw}/A_{K,s}^{\rm raw}(0)$. This rescaling leaves the eigenvectors unchanged and preserves the state dependence of the individual momentum fractions.

The transverse momentum radius is defined by the near-forward slope,
\begin{equation}
 r_{A,\perp}^2=-\frac{4(\hbar c)^2}{A_K(0)}
 \left.\frac{dA_K(-Q^2)}{dQ^2}\right|_{Q^2=0}.
 \label{eq:Aradius}
\end{equation}
For comparison with a three-dimensional form-factor slope convention, the coefficient $4$ is replaced by $6$, giving $r_A^{(3D)}=\sqrt{3/2}\,r_{A,\perp}$. This relation specifies the $A$-slope convention used in the radius table.

\subsection{Belinfante shear extraction and sector-changing operators}
\label{sec:theoryD}

The mechanical form factor requires the transverse components of Eq.~\eqref{eq:belinfante}. We extract $D_a$ from the off-diagonal shear component $T^{12}$. This choice is supported by covariant light-front analyses and perturbative benchmarks \cite{CaoEtAl2024LFEMT,CaoLiVary2023Scalar,Hu2025ShearStress}. In particular, the $T^{12}$ extraction does not require the impulse ansatz used in the corresponding $T^{+-}$ construction or an independent assumption of EMT current conservation.
The traceless transverse projection separates the $D_a$ contribution from the isotropic $\bar C_a$ term in Eq.~\eqref{eq:spinzeroGFF}. In the Drell-Yan frame, with $i,j=x,y$, define the traceless transverse projector
\begin{equation}
 \Pi_{ij}(\bm q)=\frac{q_iq_j}{Q^2}-\frac12\delta_{ij},\quad
 D_a(-Q^2)=\frac4{Q^2}\Pi_{ij}
 \langle P'|T_a^{ij}|P\rangle.
 \label{eq:Dproj}
\end{equation}
For the off-diagonal shear component $i\ne j$ this is equivalent to
\begin{equation}
 \langle P'|T_a^{12}|P\rangle=\frac12q_xq_yD_a(-Q^2),
 \qquad D_a(-Q^2)=\frac{2\mathcal M_a^{12}}{q_xq_y},
 \label{eq:D12}
\end{equation}
provided $q_xq_y\ne0$ and the relevant spurious transverse-traceless terms are absent under the stated projected-current prescription. For the numerical calculation we integrate the full transfer azimuth $0\le\phi_q<2\pi$ and compare the angular directions as a check of rotational covariance. The traceless projection is used for the azimuthal integration, avoiding division by $q_xq_y$ at its zeros. At zero transfer, we use the analytic derivative
\begin{equation}
 D_a(0)=2\left.\frac{\partial^2\mathcal M_a^{12}(q_x,q_y)}
 {\partial q_x\partial q_y}\right|_{\bm q=0}.
 \label{eq:Dorigin}
\end{equation}
The derivatives act directly on the HO translation moments and polynomial momentum insertions.

We first expand the QCD Belinfante tensor and then project it onto the retained Fock sectors. The resulting shear matrix element contains diagonal terms from the valence and one-gluon states, both directions of the Fock-changing transition, and constrained-field contact operators:
\begin{align}
 \mathcal M^{12}_B={}&\mathcal M_{u,22}^{12}
 +\mathcal M_{\bar s,22}^{12}
 +\mathcal M_{u,33}^{12}
 +\mathcal M_{\bar s,33}^{12}
 +\mathcal M_{g,33}^{12}+\sum_{a=u,\bar s,g}\bigl[Z_a^{12}(\bm q)+Z_a^{12}(-\bm q)^*\bigr]
 +\sum_{a=u,\bar s,g}\mathcal M_{a,\mathrm{ct}}^{12}.
 \label{eq:DcurrentSum}
\end{align}
The $Z_a^{12}$ denote one orientation of the $2\leftrightarrow3$ current; the second term supplies the Hermitian-conjugate transition. The diagonal quark and antiquark kernels contain kinetic terms and helicity-dependent insertions. With $x_a$ the active parton fraction, $\lambda_a=\pm1$ twice its physical helicity, and $\bar{\bm k}_a=(\bm k_a'+\bm k_a)/2$, a representative transverse shear kernel is
\begin{equation}
 \mathcal K_{a,\lambda}^{12}=\frac1{x_a}
 \left[2\bar k_{a,x}\bar k_{a,y}
 +\frac{i\lambda_a}{2}(q_x\bar k_{a,x}-q_y\bar k_{a,y})\right].
 \label{eq:qdiagMain}
\end{equation}
The diagonal $3\to3$ gluon contribution is obtained from
the field-strength part of the EMT. The need to retain gluonic EMT contributions is also seen in dressed-quark light-front calculations and lattice determinations of gluon form factors \cite{More2023GluonDressed,Pefkou2022Hadron,ShanahanDetmold2019Gluon}. Its transverse shear insertion includes the bilinear gluon momenta, helicity-dependent terms, and a term proportional to $q_xq_y$ times the wave-function overlap without a momentum insertion. All three terms are retained in the gluon shear-current contraction.

The one-body overlap does not capture the full Fock sector changing $T^{12}$ matrix element. Interaction-current contributions have also been studied in light-front Hamiltonian EMT calculations \cite{CaoLiVary2023Scalar,More2023GluonDressed}. For the quark and antiquark contributions, the interaction current contains both the explicit transverse field in $\overleftrightarrow D$ and the constrained-fermion contributions from eliminating $\psi_-$. In the gluon sector the eliminated $A^-$ field produces interactions involving quark and antiquark color currents. The signed $R_{1,\epsilon}$ denominators appear in these insertions as specified by the QCD constraints. The $g^2$ contact terms from the constrained-field operators also appear with opposite contributions in the quark and gluon sectors, satisfying the route identities
\begin{equation}
 \mathcal M_{u,\mathrm{ct}}^{12}=-\mathcal M_{g,\mathrm{ct},u}^{12},\qquad
 \mathcal M_{\bar s,\mathrm{ct}}^{12}=-\mathcal M_{g,\mathrm{ct},\bar s}^{12}.
 \label{eq:contactsCancelApp}
\end{equation}
These identities apply to the contact pieces of the projected current. The total form factor is
\begin{equation}
 D_K(t)=D_u(t)+D_{\bar s}(t)+D_g(t),
 \label{eq:DpartonSum}
\end{equation}
under the projected-current prescription. We evaluate the forward derivatives and finite-transfer contractions in the same basis, retaining the separate gluon contribution.  The separation of the force contributions into quark and gluon parts depends on the operator definition \cite{PolyakovSon2018,HattaRajanTanaka2018}.

\subsection{Transverse mechanical observables}
\label{sec:theoryMechanics}

We calculate $D_a(-Q^2)$ directly at 21 momentum-transfer
points, $Q^2=0,0.1,\ldots,2.0\,\mathrm{GeV}^2$, for
$a=u,\bar s,g$. Each value is evaluated using the corresponding
eigenstate and the projected Belinfante operator, and the
total form factor $D_K$ is obtained by summing the three
contributions. The pressure and shear distributions are obtained from
Fourier-Bessel transforms of the $D$ form factor over the
full spacelike momentum-transfer range. We therefore need
a continuation beyond the interval calculated directly. Related meson analyses use momentum-transfer parameterizations with their own ultraviolet assumptions \cite{Broniowski2008Pion,Broniowski2024MesonDominance, XuDing2024Kaon}. We use a shape-preserving piecewise cubic Hermite interpolation (PCHIP) on the direct interval and a fitted rational function for the ultraviolet tail.

For $\tau=Q^2\ge0$ and $\tau_0=2\,\mathrm{GeV}^2$, we define Ansatz 1 by $n=5$ and Ansatz 2 by $n=6$ in
\begin{equation}
 F_{a,n}(\tau)=\frac{1+B_{a,n}\tau+C_{a,n}\tau^2}{(1+\tau/A_{a,n})^n},
 \quad A_{a,n}>0,
 \label{eq:Fn}
\end{equation}
where the parameters are fitted to the central values at the 20 nonzero momentum-transfer points, with $D_a(0)$ fixed. The continued form factor for eigenstate $s$ is
\begin{equation}
 D_{a,s}^{(n)}(-\tau)=\begin{cases}
 D_{a,s}^{\rm PCHIP}(-\tau),&0\le \tau\le \tau_0,\\
 D_{a,s}(-\tau_0)F_{a,n}(\tau)/F_{a,n}(\tau_0),&\tau>\tau_0.
 \end{cases}
 \label{eq:piecewiseD}
\end{equation}
The matching is continuous, but its derivative is not constrained to match. At the same $n$, the fitted ultraviolet \emph{shape} is fixed across the ten eigenstates, while each state retains its own direct curve and matching value $D_{a,s}(-\tau_0)$. For large $Q^2$ the two ansätze behave as $(Q^2)^{-3}$ and $(Q^2)^{-4}$ when the highest numerator coefficient is nonzero. These are model assumptions applied only above $\tau_0$, not calculated QCD asymptotics. Their difference estimates the sensitivity to these two ultraviolet completions.

For each eigenstate, we define the transverse pressure and shear distributions using its calibrated mass $M_K$ and $\hbar c=0.1973269804\,\mathrm{GeV\,fm}$, where $b_\perp=|\bm b_\perp|$, 
\begin{align}
 p_K(b_\perp)&=-\frac1{32\pi M_K(\hbar c)^2}
 \int_0^\infty\!d\tau\tau J_0\left(\frac{b_\perp\sqrt \tau}{\hbar c}\right)D_K(-\tau),
 \label{eq:pressureInt}\\
 s_K(b_\perp)&=-\frac1{16\pi M_K(\hbar c)^2}
 \int_0^\infty\!d\tau\tau J_2\left(\frac{b_\perp\sqrt \tau}{\hbar c}\right)D_K(-\tau).
 \label{eq:shearInt}
\end{align}
These two-dimensional distributions are defined by the $D$ form factor in the light-front convention \cite{FreeseMiller2021,PolyakovSchweitzer2018,Miller2007Transverse,LiWangVary2024EMT}. The principal transverse stresses and radially weighted distributions are
\begin{equation}
 \begin{aligned}
 \sigma_{r,K}(b_\perp)&=p_K(b_\perp)+\tfrac12s_K(b_\perp), \qquad
 \sigma_{t,K}(b_\perp)=p_K(b_\perp)-\tfrac12s_K(b_\perp),
 \end{aligned}
 \label{eq:normalStresses}
\end{equation}
\begin{equation}
 \begin{aligned}
 \mathcal P_K(b_\perp)&=2\pi b_\perp\,p_K(b_\perp), \qquad
 \mathcal S_K(b_\perp)=2\pi b_\perp\,s_K(b_\perp).
 \end{aligned}
 \label{eq:weightedStresses}
\end{equation}
For comparisons with plots of $2\pi b_\perp P^+p^{\rm LF}$, we convert to a common stress convention. Differences between stress-density conventions and their associated conservation relations have been discussed in light-front and effective hadron models \cite{CaoLiVary2023Scalar,LiWangVary2024EMT}. The transverse equilibrium identity,
\begin{equation}
 \frac{d\sigma_{r,K}(b_\perp)}{db_\perp}+\frac{s_K(b_\perp)}{b_\perp}=0,
 \label{eq:localEquilibrium}
\end{equation}
follows from this decomposition and provides a check of the Fourier-Bessel calculation.

We evaluate the rational tail analytically and the direct interval by deterministic quadrature. Subtracting the rational contribution on the direct interval avoids double counting and removes the need for a numerical high-transfer cutoff. The transform identities and subtraction formula are given in Appendix~\ref{app:transforms}.

The two-dimensional mechanical mean-square radius is
\begin{equation}
 \langle b_\perp^2\rangle_{\rm mech}^{(n)}
 =4(\hbar c)^2\frac{D_K(0)}{\displaystyle\int_0^\infty d\tau\,D_K^{(n)}(-\tau)},
 \label{eq:mechradius}
\end{equation}
which should not be confused with the slope-derived mass radius or a three-dimensional radius. Its continuation-family variation and the ultraviolet sensitivity of the central pressure are reported explicitly.

For the three-dimensional mechanical convention quoted in Table~\ref{tab:kaon_radii_comparison}, we use
\begin{equation}
 \bigl(r_{\rm mech}^{(3D)}\bigr)^2
 =6(\hbar c)^2\frac{D_K(0)}{\displaystyle\int_0^\infty d\tau\,D_K^{(n)}(-\tau)}
 =\frac32\langle b_\perp^2\rangle_{\rm mech}^{(n)}.
 \label{eq:mechradius3D}
\end{equation}
The pressure-form-factor slope radius is defined separately by
\begin{equation}
 \bigl(r_{\theta_1}^{(3D)}\bigr)^2
 =-\frac{6(\hbar c)^2}{\theta_1(0)}
 \left.\frac{d\theta_1(-Q^2)}{dQ^2}\right|_{Q^2=0}.
 \label{eq:pressureSlopeRadius}
\end{equation}
Here $\theta_1$ denotes the pressure form factor in the comparison convention. A constant normalization or sign relating it to $D_K$ cancels in the normalized slope. Unlike the mechanical radius, this radius depends on the behavior near $Q^2=0$ rather than an integral over all momentum transfers.

\begin{figure*}[!tbp]
    \centering
    \paperfigure{0.48\textwidth}{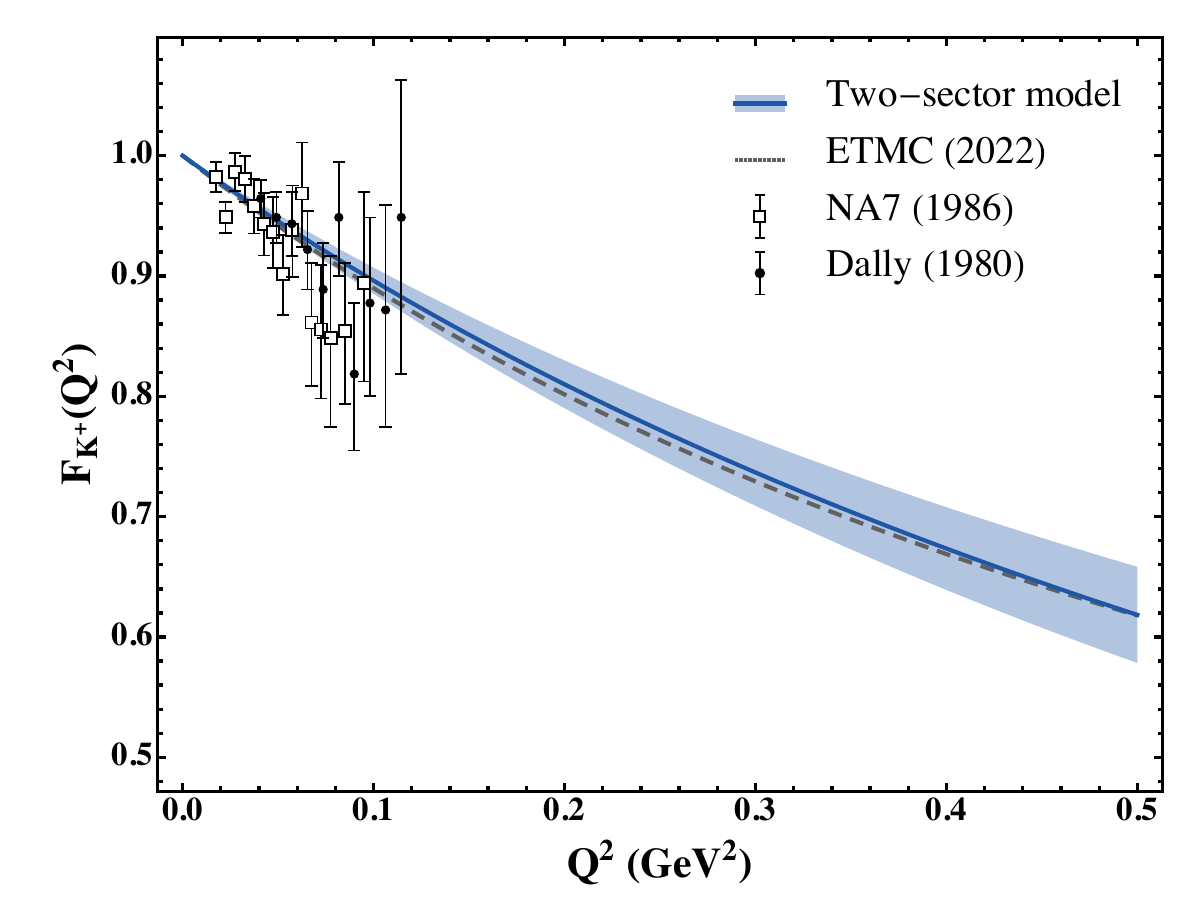}
    \hfill
    \paperfigure{0.48\textwidth}{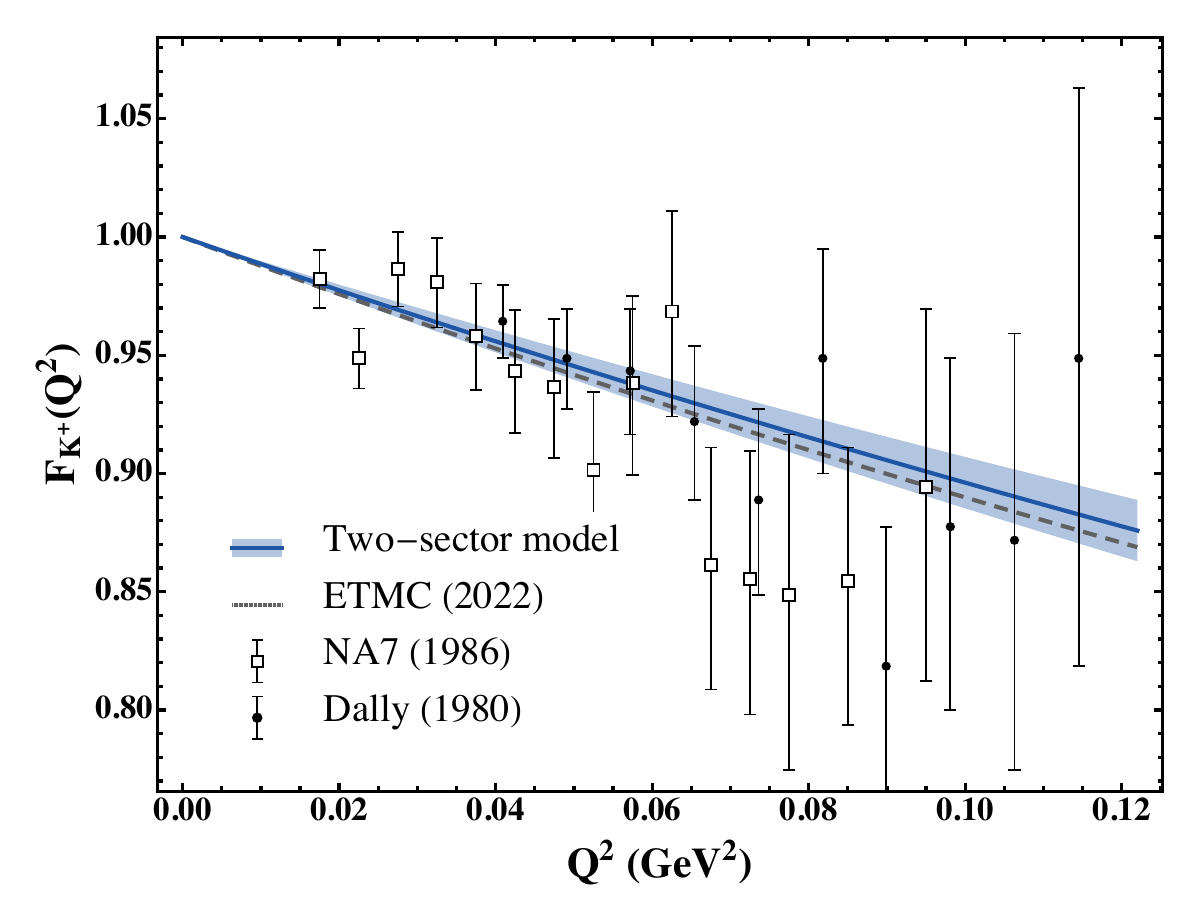}

    \caption{Charged-kaon electromagnetic form factor over
    $0\le Q^2\le0.5\,\mathrm{GeV}^2$ (left), with an expanded
    view of the low-$Q^2$ region (right). Squares and circles
    denote the NA7 and Dally data points
    \cite{Amendolia1986Kaon,Dally1980Kaon}; the dashed curve
    shows the ETMC reference result~\cite{alexandrou2022scalar}.
    The band represents the combined mass and basis sensitivity.}
    \label{fig:kaon_em_form_factor}
\end{figure*}

\section{Numerical results and discussion}
\label{sec:finalResults}

\subsection{Numerical setup and uncertainty estimates}
\label{sec:parameters}

We use $\kappa=0.65\,\mathrm{GeV}$, $\lambda_V=6\,\mathrm{GeV}$, $m_g=0.60\,\mathrm{GeV}$, and $\epsilon=0.10$. A basis scan at the central quark masses gives $(c_3,s_\beta)=(1.05,1.25)$.

The effective quark-mass parameters are taken from lattice
calculations. For the $u$ quark, we use the chiral-limit infrared mass function $M(p^2=0)=298\pm8\pm30\,\mathrm{MeV}$ obtained by
\cite{SkullerudWilliams2001}. For the strange quark, we
use the kinetic mass $m_s=554(19)(6)(8)\,\mathrm{MeV}$ extracted from the $s\bar s$ Bethe-Salpeter amplitude \cite{KawanaiSasaki2015}. We combine the quoted uncertainties in quadrature and use $m_u=(0.298\pm0.03105)\,\mathrm{GeV}, m_s=(0.554\pm0.02147)\,\mathrm{GeV}$.
These two lattice quantities have different definitions
and are used as effective mass parameters in our
Hamiltonian. In the sensitivity analysis, their
variations are treated independently. We denote the
lower and upper variations by $u^-$ and $u^+$ for the
$u$ quark, and by $s^-$ and $s^+$ for the strange quark,
respectively. The central calculation uses the reference basis,
ed
$(N_{\max}^{(2)},N_{\max}^{(3)})=(12,16)$. We also
repeat the calculation with the reduced basis,
$(N_{\max}^{(2)},N_{\max}^{(3)})=(6,8)$, to estimate
the sensitivity to the basis truncation. For each basis
and each mass choice, the coupling $g$ is recalibrated
to reproduce the physical kaon mass,
$M_{K^+}=0.493677\,\mathrm{GeV}$, while all other model
parameters are kept fixed. For the central quark masses, the calibrated coupling is $g=1.0525720$ in the reference basis and
$g=0.9885420$ in the reduced basis. For each mass-varied
state, the coupling is recalibrated independently to
reproduce the physical kaon mass. 

The partonic matrix elements are evaluated at the effective
model scale $\mu_0=0.580\,\mathrm{GeV}$, determined by
evolving the forward singlet-quark and gluon momentum
fractions at NNLO in the $\overline{\mathrm{MS}}$ scheme,
with charm-threshold matching at
$\overline{m}_c(\overline{m}_c)=1.27\,\mathrm{GeV}$, to match
$A_g(0,\mu)/A_\Sigma(0,\mu)$ to the physical-point
lattice-QCD ratio at $\mu=2\,\mathrm{GeV}$~\cite{Alexandrou2025PionKaonMom}.

For an observable $\mathcal O$, we estimate the
sensitivity to the quark-mass variations and basis
truncation as
\begin{equation}
\begin{aligned}
\delta_{\rm mass}\mathcal O
=\frac12\Bigl[
(\mathcal O_{u+}^{\rm ref}-\mathcal O_{u-}^{\rm ref})^2
+(\mathcal O_{s+}^{\rm ref}-\mathcal O_{s-}^{\rm ref})^2
\Bigr]^{1/2},\,
\delta_{\rm basis}\mathcal O
=|\mathcal O_0^{\rm ref}-\mathcal O_0^{\rm low}|, \,
\delta_{\rm MB}\mathcal O
=\left[
(\delta_{\rm mass}\mathcal O)^2+
(\delta_{\rm basis}\mathcal O)^2
\right]^{1/2},
\end{aligned}
\label{eq:errors}
\end{equation}
where $\mathrm{ref}$ and $\mathrm{low}$ denote the
reference and reduced bases, respectively, and $0$
denotes the central quark masses. The coupling
recalibration is included in each mass-varied state.
For total observables, the partonic contributions are
summed separately for each state before evaluating
the sensitivity, preserving their correlations.
The uncertainty bands show $\delta_{\rm MB}$ and
represent model sensitivity rather than statistical
confidence intervals. The difference between the
two ultraviolet continuations is reported separately.

\begin{figure*}[t]
    \centering

    \begin{minipage}[t]{0.48\textwidth}
        \centering
        \paperfigure{0.98\linewidth}
        {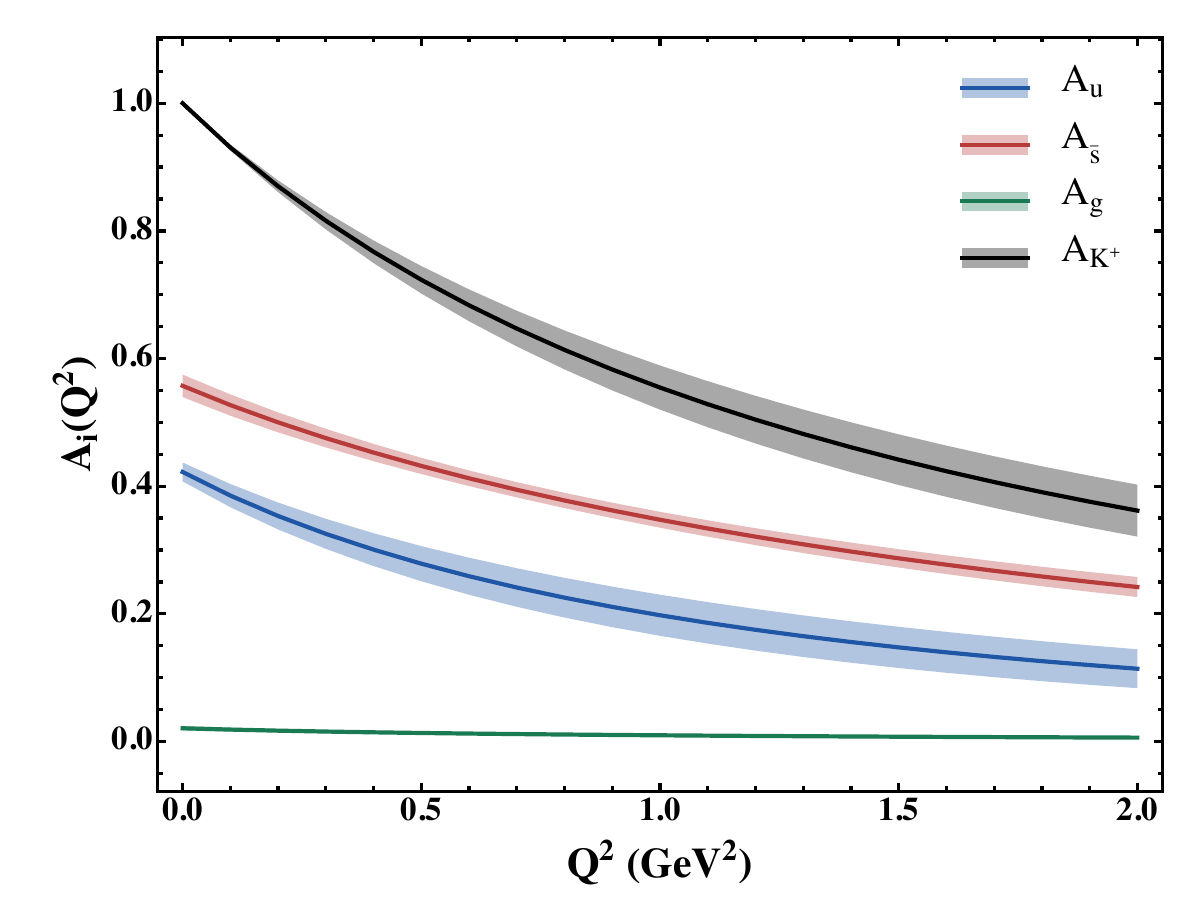}

        \caption{Momentum gravitational form factors $A_a(Q^2)$
        of the charged kaon. Blue, red, green, and black colour denote
        the $u$-quark, strange-antiquark, gluon, and total
        contributions, respectively. The bands show the combined
        mass and basis sensitivity.}
        \label{fig:kaon_A_form_factors}
    \end{minipage}
    \hfill
    \begin{minipage}[t]{0.48\textwidth}
        \centering
        \paperfigure{0.98\linewidth}
        {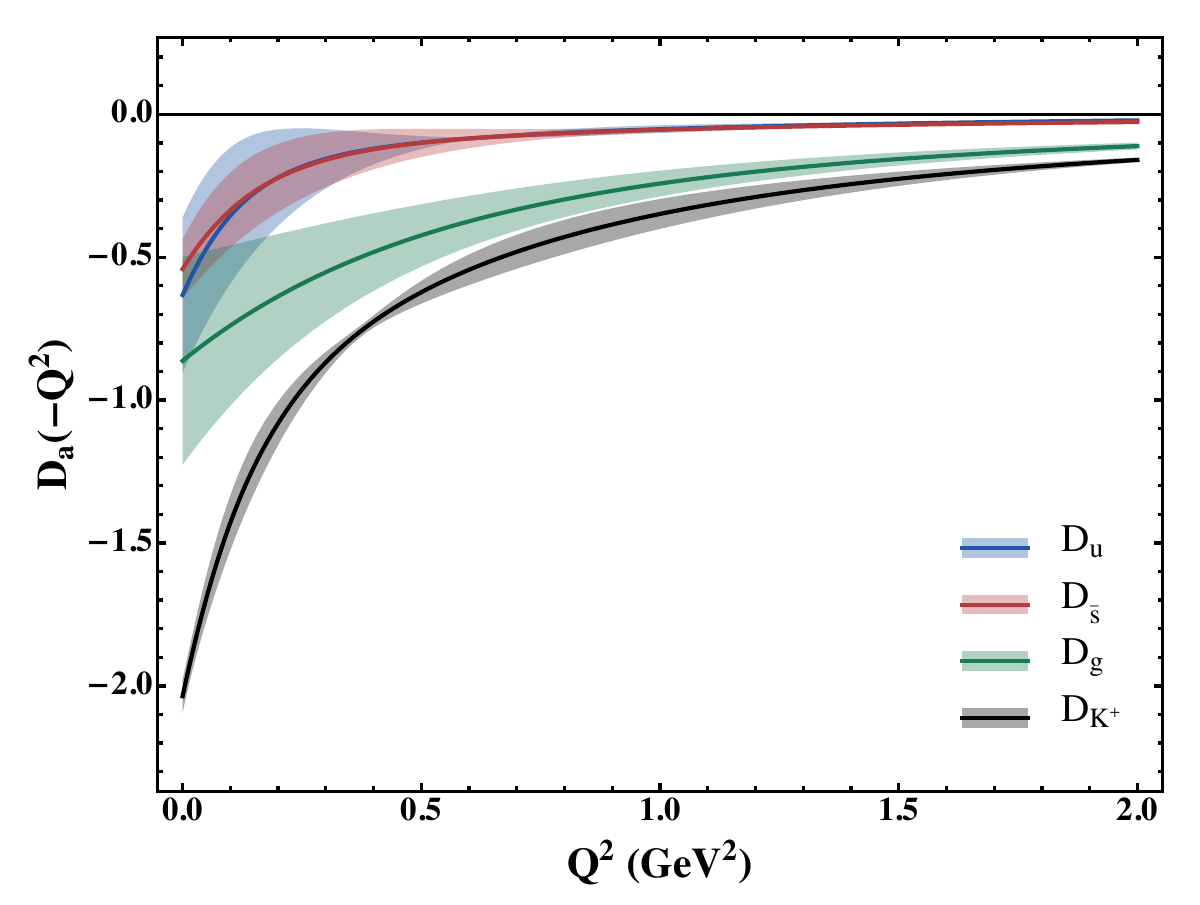}

        \caption{The $D$-term gravitational form factors
        $D_a(-Q^2)$ for $a=u,\bar{s},g,K^+$ are plotted over
        the range $0\le Q^2\le2.0\,\mathrm{GeV}^2$. The blue,
        red, green, and black colour curves represent the $u$-quark,
        strange-antiquark, gluon, and total contributions,
        respectively. The bands show the combined mass and
        basis sensitivity.}
        \label{fig:kaon_D_form_factors}
    \end{minipage}
\end{figure*}

\subsection{Electromagnetic and gravitational form factors}
\label{sec:resultsFormFactors}
\label{sec:resultsEM}
\label{sec:resultsA}
\label{sec:resultsD}

We first consider the electromagnetic form factor of the
charged kaon. Figure~\ref{fig:kaon_em_form_factor} compares
our result for $F_{K^+}$ with the NA7 and Dally measurements
\cite{Amendolia1986Kaon,Dally1980Kaon} and the ETMC
reference curve~\cite{alexandrou2022scalar}. At low $Q^2$, the calculated form factor
follows the decrease observed in the experimental data
and lies slightly above the ETMC curve over most of the
displayed interval. The slope at the origin gives the
charge radius
$r_{\rm ch}=0.5178\pm0.0278\,\mathrm{fm}$.

The forward values of the momentum gravitational form
factors are
\begin{equation}
\begin{aligned}
A_u(0)&=0.42224\pm0.01481,\quad
A_{\bar s}(0)=0.55723\pm0.01758, \quad
A_g(0)=0.02053\pm0.00356,
\quad A_K(0)=1.
\end{aligned}
\label{eq:AforwardResults}
\end{equation}
The strange antiquark carries the largest fraction of
the kaon momentum, while the gluon contribution is
$2.05\%$. This should be distinguished from the
three-body Fock-sector probability, $P_3\simeq9.67\%$,
which measures the probability of finding the
$|u\bar s g\rangle$ component rather than the fraction
of momentum carried by the gluon. Related pion
calculations have also resolved an explicit gluon
contribution \cite{Lan2025PionGluonFock}.
The state-specific normalization described in
Sec.~\ref{sec:theoryA} corrects the small quadrature
deviation $A_K^{\rm raw}(0)=0.999986954$ and enforces
the momentum sum rule for each state.

Figure~\ref{fig:kaon_A_form_factors} shows the momentum
transfer dependence of the individual contributions.
The strange-antiquark form factor decreases more slowly
than the $u$-quark contribution and remains dominant.
The gluon contribution is small over the displayed
range. The individual partonic curves are shown at $\mu_0 = 0.580~\mathrm{GeV}$; comparing them directly with lattice-QCD results at $\mu = 2~\mathrm{GeV}$~\cite{Alexandrou2025PionKaonMom,Alexandrou2020Momentum} requires singlet evolution and quark-gluon operator mixing, whereas the total form factor $A_K(Q^2)$ is scale invariant.
The total form factor can also be compared with
valence-sector BLFQ, continuum, and dispersive
calculations \cite{Sain2026KaonBLFQ,XuDing2024Kaon,
CaoGuo2025Dispersive}.

The forward values of the Belinfante $D$ form factors
are
\begin{equation}
\begin{aligned}
D_u(0)=-0.63224, \quad
D_{\bar s}(0)=-0.54131, \quad
D_g(0)=-0.86255,  \quad
D_K(0)=-2.03610.
\end{aligned}
\label{eq:DforwardValues}
\end{equation}
The $u$-quark, strange-antiquark, and gluon contributions
account for $31.1\%$, $26.6\%$, and $42.4\%$ of
$|D_K(0)|$, respectively. Thus, although the gluon
carries only $2.05\%$ of the longitudinal momentum, its
contribution to the total $D$ term is substantial.
This difference reflects the structure of the transverse
Belinfante current, which contains interaction,
Fock-sector-changing, and constrained-field
contributions that are absent from the diagonal
$T^{++}$ overlap.

As shown in Fig.~\ref{fig:kaon_D_form_factors}, the
partonic $D$ form factors are negative at the origin
and approach zero as $Q^2$ increases. The quark
contributions decrease more rapidly in magnitude than
the gluon contribution, which becomes dominant at
larger momentum transfer. These form factors are
calculated directly from the current matrix elements
and are not affected by the ultraviolet continuation
introduced later for the mechanical distributions.
We check the current contractions through their
azimuthal dependence and the contact identities
in Eq.~\eqref{eq:contactsCancelApp}.

The explicit gluon sector therefore extends the
valence-sector BLFQ description by including both
$D_g$ and the Fock-sector-changing current
contributions \cite{Sain2026KaonBLFQ}. Comparisons
with dressed-vertex calculations
\cite{XuDing2024Kaon,LiuAbuki2026KaonNJL} and
covariant light-front analyses
\cite{CaoEtAl2024LFEMT} depend on the Hamiltonian
and current prescriptions used. Gluon gravitational
form factors have also been investigated in proton
spectator models \cite{Sain2025Gluon}.

\begin{figure*}[!tbp]
    \centering

    \begin{subfigure}[t]{0.48\textwidth}
        \centering
        \paperfigure{\linewidth}{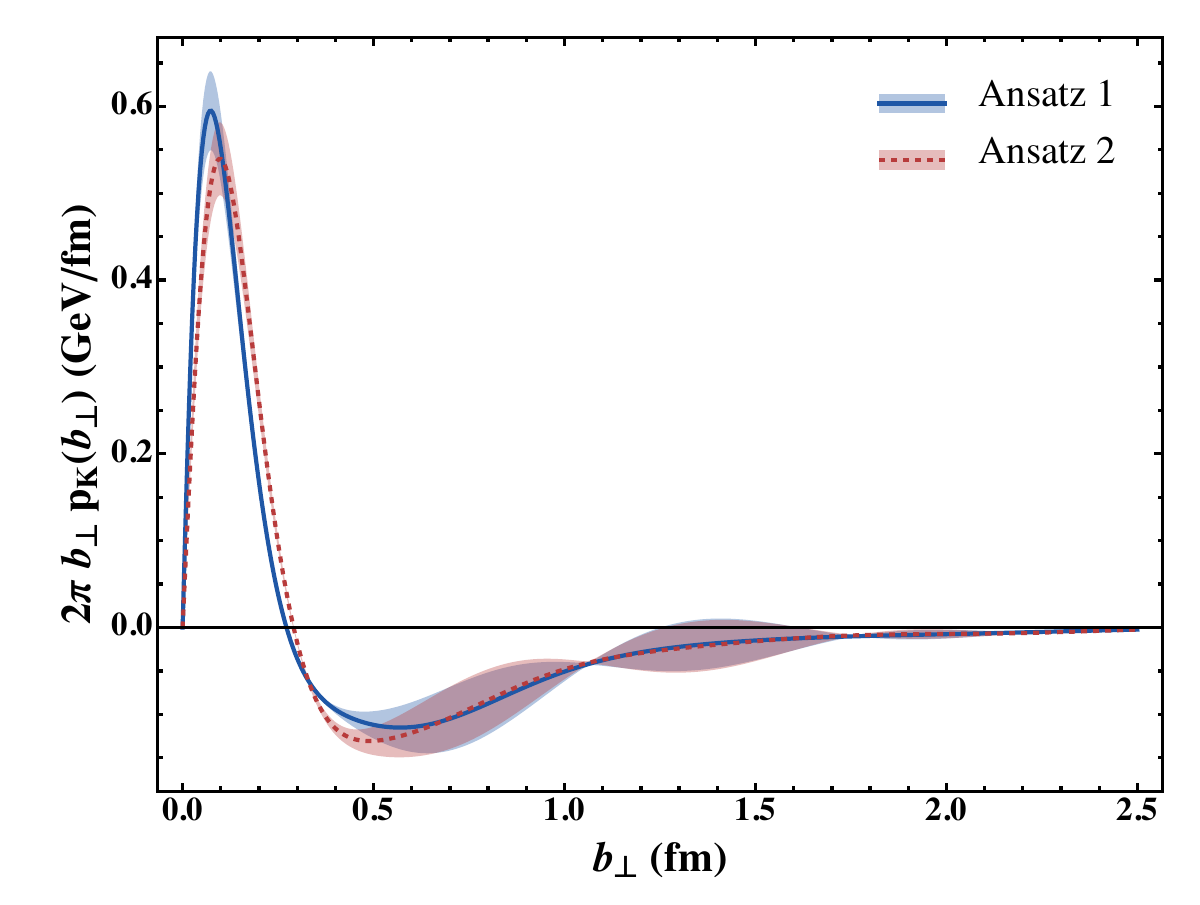}
        \caption{}
        \label{fig:kaon_pressure_comparison}
    \end{subfigure}
    \hfill
    \begin{subfigure}[t]{0.48\textwidth}
        \centering
        \paperfigure{\linewidth}{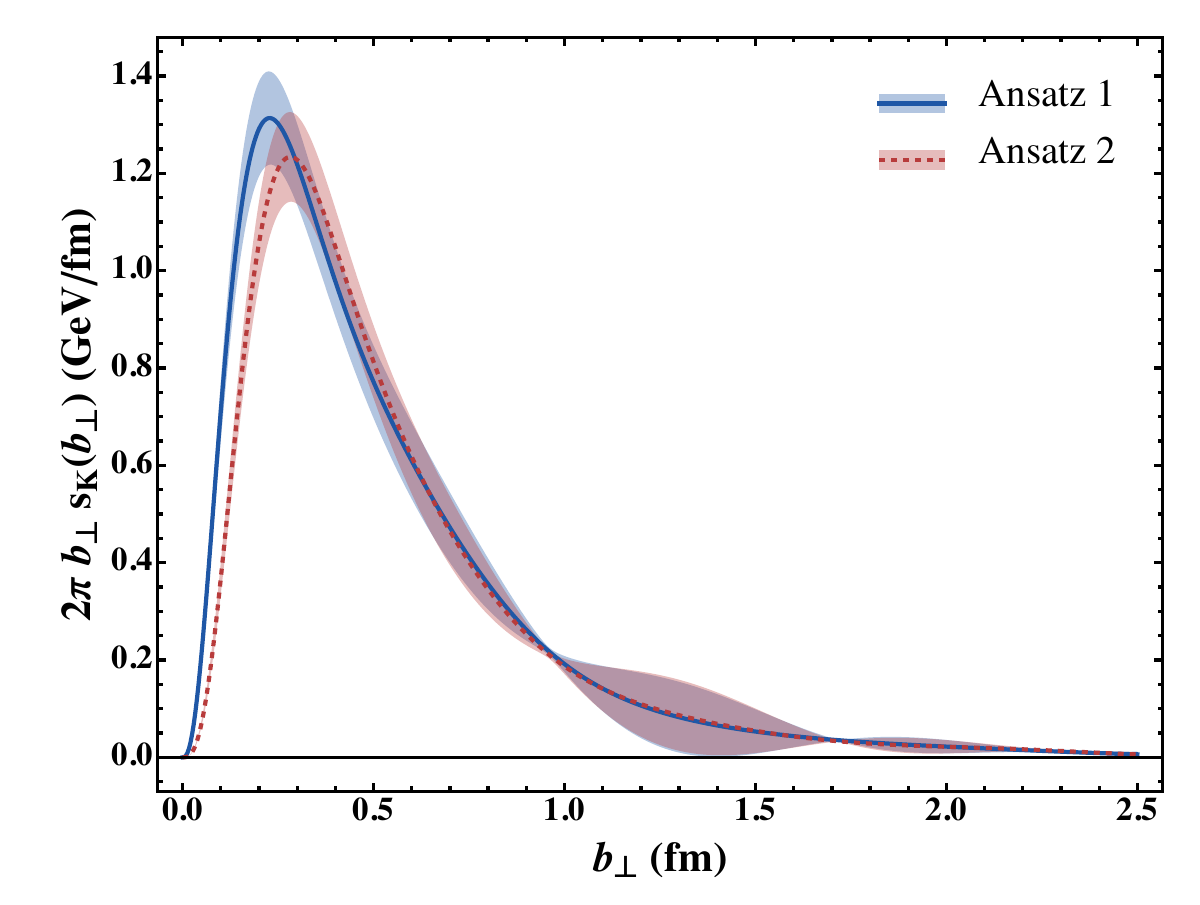}
        \caption{}
        \label{fig:kaon_shear_comparison}
    \end{subfigure}

    \vspace{0.3em}

    \begin{subfigure}[t]{0.48\textwidth}
        \centering
        \paperfigure{\linewidth}{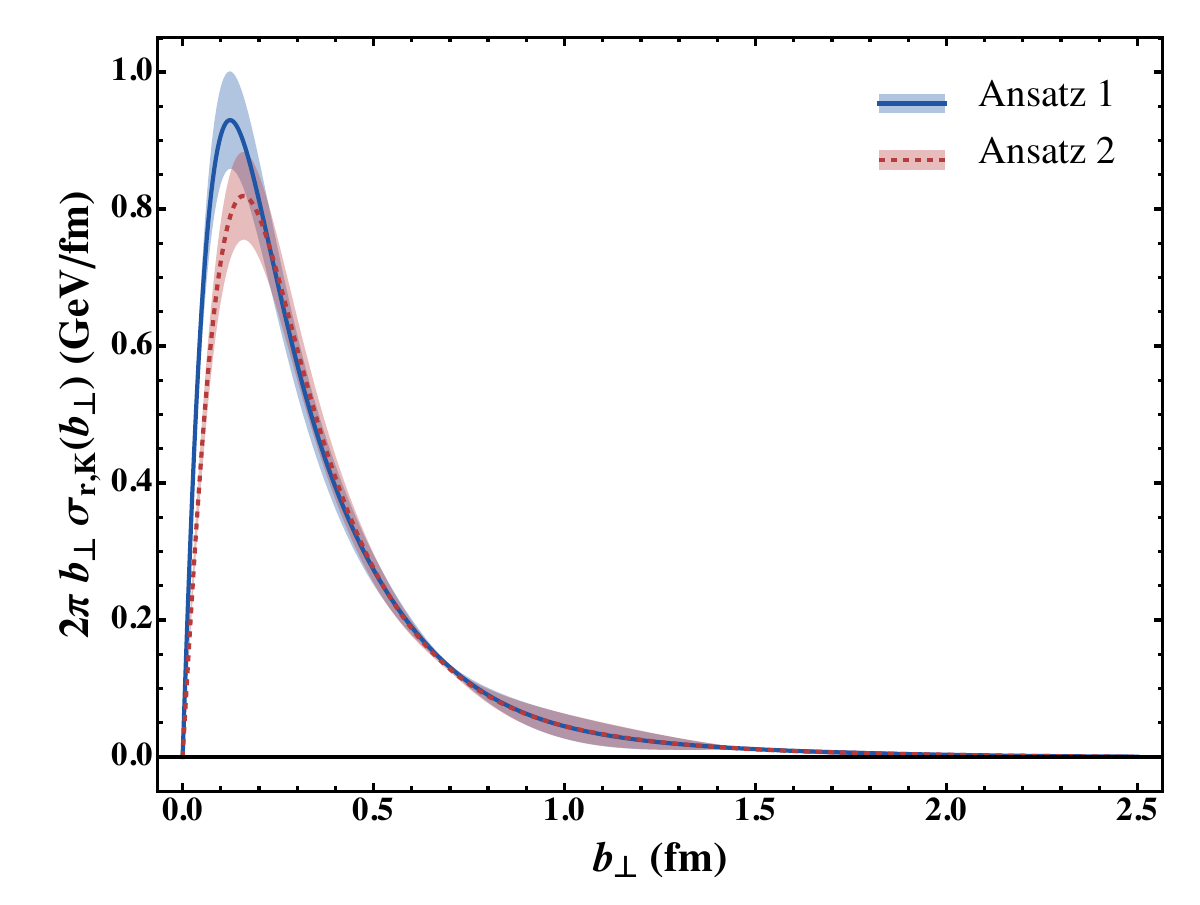}
        \caption{}
        \label{fig:kaon_radial_comparison}
    \end{subfigure}
    \hfill
    \begin{subfigure}[t]{0.48\textwidth}
        \centering
        \paperfigure{\linewidth}{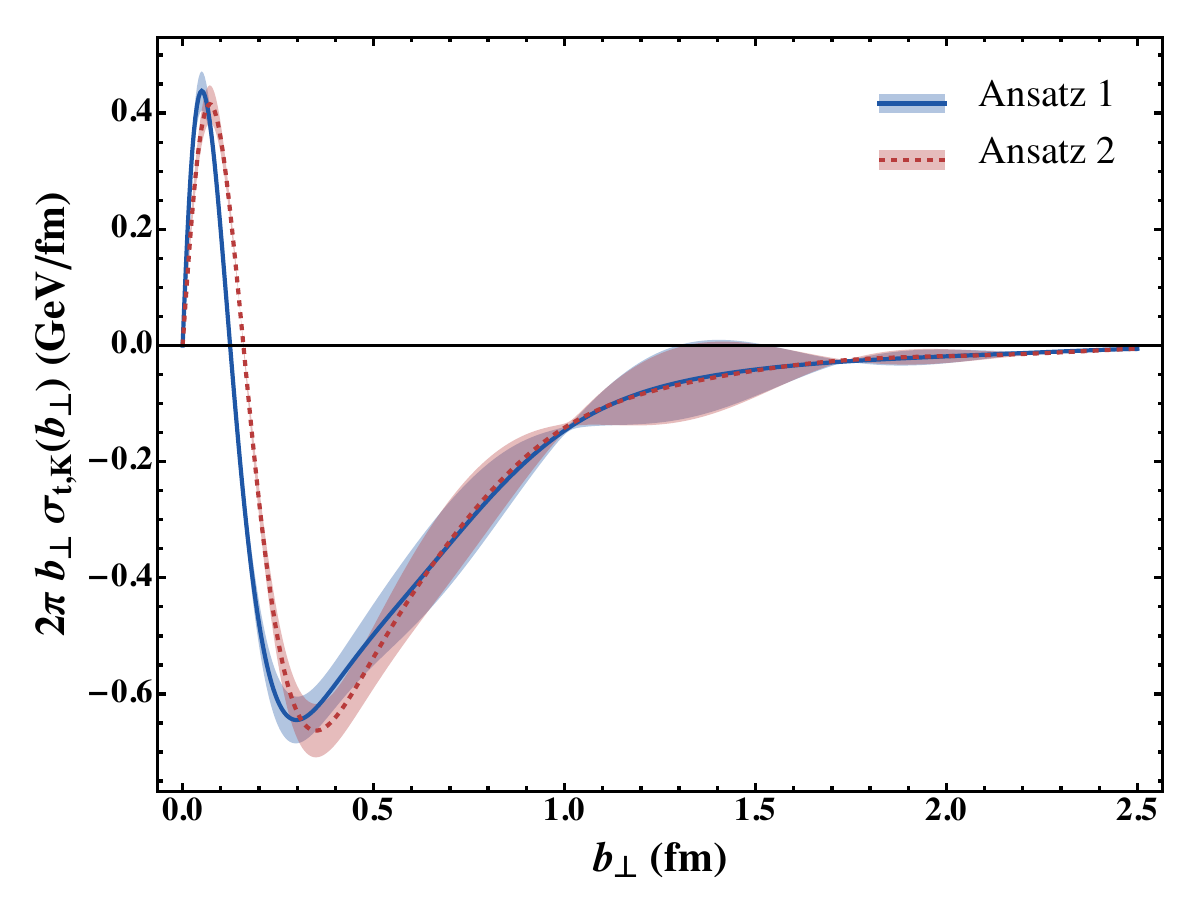}
        \caption{}
        \label{fig:kaon_tangential_comparison}
    \end{subfigure}

    \caption{Weighted transverse mechanical distributions
    plotted over $0\le b_\perp\le2.5\,\mathrm{fm}$.
    Solid blue and dotted red curves denote Ansatz 1 and
    Ansatz 2, respectively. The bands show the combined
    mass and basis sensitivity at fixed continuation shape.}
    \label{fig:kaon_mechanical_total}
\end{figure*}

\subsection{Transverse mechanical structure}
\label{sec:resultsMechanics}
\label{sec:resultsPartonic}

The transverse pressure and shear distributions require
the $D$ form factor over the full spacelike momentum-transfer
range. We calculate $D_a(-Q^2)$ directly at 21 points,
$Q^2=0,0.1,\ldots,2.0\,\mathrm{GeV}^2$, and use the
matched continuations of Eq.~\eqref{eq:piecewiseD} above
$Q^2=2\,\mathrm{GeV}^2$. We consider two rational
parameterizations, Ansatz 1 and Ansatz 2, corresponding
to $n=5$ and $n=6$, respectively, in Eq.~\eqref{eq:Fn}.
The direct results are retained throughout the calculated
interval, without an additional high-$Q^2$ cutoff or
spatial smearing.

For each partonic contribution, we fit
$D_a^{\mathrm{fit}}(-Q^2)=D_a(0)F_{a,n}(Q^2)$ to the
20 nonzero-transfer points, keeping $D_a(0)$ fixed.
The quality of the fits is characterized by the
logarithmic root-mean-square residual and the maximum
relative deviation,
\begin{align}
R_{\log}
&=
100\sqrt{\frac{1}{20}\sum_{i=1}^{20}
\left[
\ln\frac{D_a^{\mathrm{fit}}(-Q_i^2)}
         {D_a^{\mathrm{calc}}(-Q_i^2)}
\right]^2},
\\
\delta_{\max}
&=
100\max_i
\left|
\frac{D_a^{\mathrm{fit}}(-Q_i^2)}
     {D_a^{\mathrm{calc}}(-Q_i^2)}-1
\right|.
\end{align}
Both quantities are expressed in percent. The fitted
coefficients and residuals are listed in
Table~\ref{tab:kaon_pade_fit}. Ansatz 1 gives smaller
residuals in all three partonic channels. Both
parameterizations reproduce the calculated gluon
form factor to better than $0.2\%$.

\begin{table*}[!htbp]
\centering
\caption{Fitted coefficients and residuals for the partonic
$D$ form factors. The coefficient sensitivities combine
mass and basis variations obtained by refitting the
corresponding states and are not statistical errors.
The residuals refer to the central-state fits.}
\label{tab:kaon_pade_fit}

\footnotesize
\setlength{\tabcolsep}{6pt}
\renewcommand{\arraystretch}{1.12}

\begin{ruledtabular}
\begin{tabular}{llccccc}
Ansatz & Parton
& $A_{a,n}$ ($\mathrm{GeV}^2$)
& $B_{a,n}$ ($\mathrm{GeV}^{-2}$)
& $C_{a,n}$ ($\mathrm{GeV}^{-4}$)
& $R_{\log}$ (\%)
& $\delta_{\max}$ (\%)\\
\hline
1 & $u$
& $0.8347\pm0.6858$
& $-0.6641\pm0.6489$
& $3.8983\pm3.0905$
& $0.456$ & $1.472$\\

1 & $\bar s$
& $1.1264\pm0.7101$
& $-0.7633\pm0.0396$
& $2.1645\pm1.5851$
& $0.580$ & $1.227$\\

1 & $g$
& $3.3837\pm4.7201$
& $-0.1084\pm0.2672$
& $0.1329\pm0.0533$
& $0.052$ & $0.101$\\
\hline
2 & $u$
& $1.1899\pm0.7470$
& $-1.1596\pm0.2428$
& $3.3954\pm2.5143$
& $0.983$ & $1.683$\\

2 & $\bar s$
& $1.5335\pm0.7870$
& $-1.0558\pm0.2309$
& $2.0980\pm1.4592$
& $1.298$ & $2.660$\\

2 & $g$
& $4.1271\pm6.0822$
& $-0.1268\pm0.2683$
& $0.1584\pm0.0726$
& $0.069$ & $0.159$\\
\end{tabular}
\end{ruledtabular}
\end{table*}

In constructing the mechanical uncertainty bands, we retain
the direct form-factor values and the matching value of
each mass-varied and basis-varied state, while keeping
the central-state ultraviolet shape fixed. The coefficient
sensitivities listed in Table~\ref{tab:kaon_pade_fit} are
therefore not added as separate uncertainties. The bands
reflect the mass and basis variations of the underlying
eigenstates.

Figure~\ref{fig:kaon_mechanical_total} shows the
$2\pi b_\perp$-weighted transverse pressure, shear,
radial stress, and tangential stress for the two
continuations. Both give a positive-pressure region
at small transverse distances, followed by a broader
negative-pressure region. The shear and radial stress
remain positive in the inner region, whereas the
tangential stress changes sign at a smaller distance
than the pressure and develops a negative minimum.
Ansatz 1 gives higher peaks in the weighted pressure,
shear, and radial-stress distributions. At large
$b_\perp$, the distributions become small and the
two continuations approach one another.

\begin{figure*}[!tbp]
    \centering

    \begin{subfigure}[t]{0.48\textwidth}
        \centering
        \paperfigure{\linewidth}{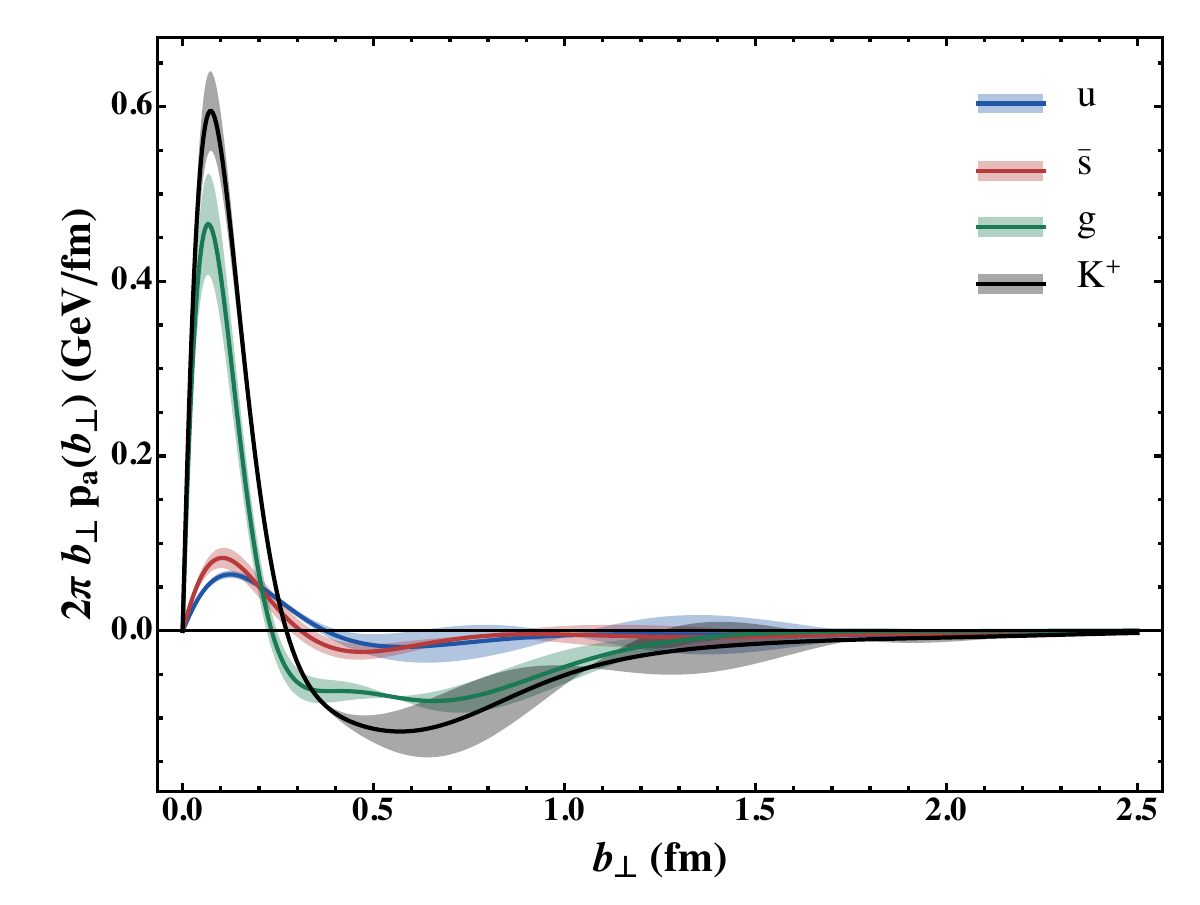}
        \caption{}
        \label{fig:partonic_pressure}
    \end{subfigure}
    \hfill
    \begin{subfigure}[t]{0.48\textwidth}
        \centering
        \paperfigure{\linewidth}{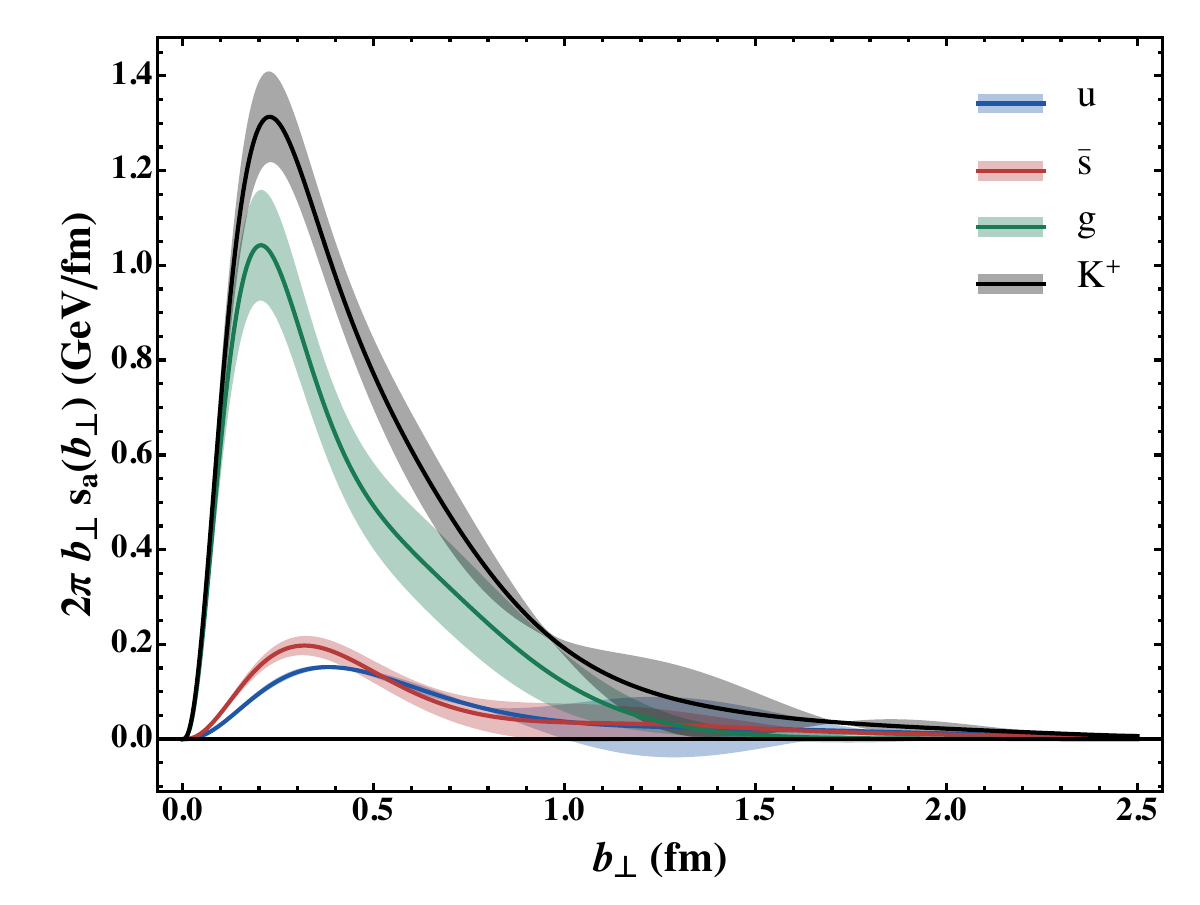}
        \caption{}
        \label{fig:partonic_shear}
    \end{subfigure}

    \vspace{0.3em}

    \begin{subfigure}[t]{0.48\textwidth}
        \centering
        \paperfigure{\linewidth}{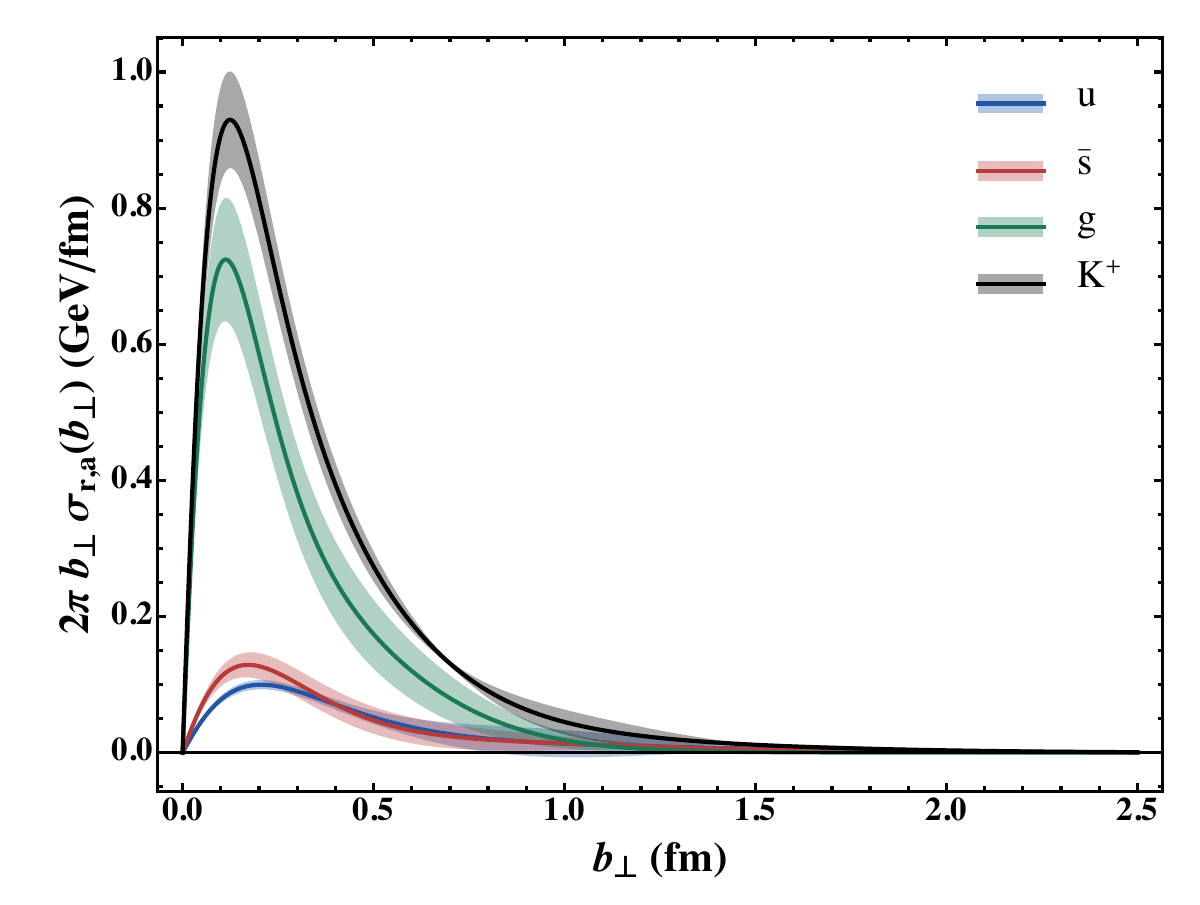}
        \caption{}
        \label{fig:partonic_radial}
    \end{subfigure}
    \hfill
    \begin{subfigure}[t]{0.48\textwidth}
        \centering
        \paperfigure{\linewidth}{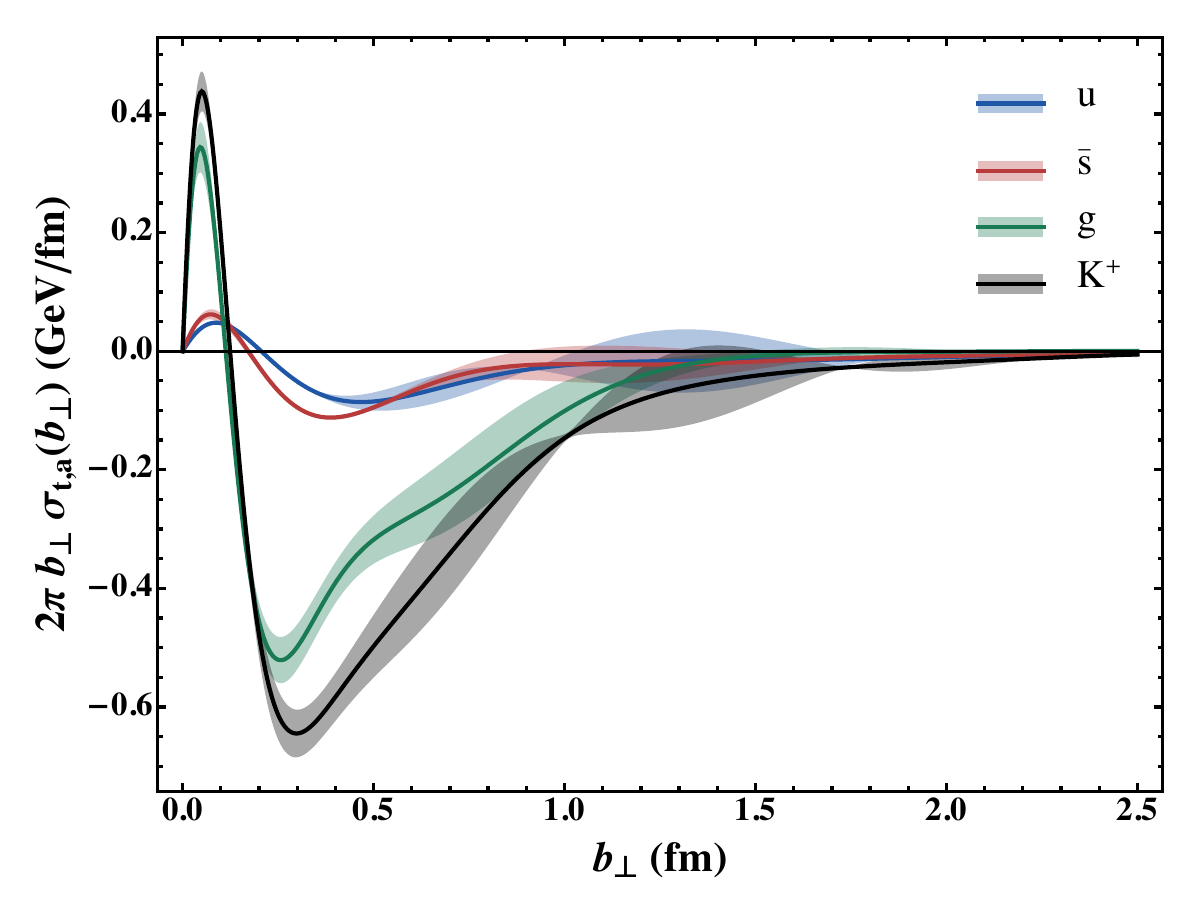}
        \caption{}
        \label{fig:partonic_tangential}
    \end{subfigure}

    \caption{Partonic contributions to the weighted transverse
    pressure (a), shear (b), radial stress (c), and tangential
    stress (d) for Ansatz 1, plotted over
    $0\le b_\perp\le2.5\,\mathrm{fm}$. Blue, red, green,
    and black colour curves denote the $u$-quark, strange-antiquark,
    gluon, and total contributions, respectively. The bands
    show the combined mass and basis sensitivity.}
    \label{fig:kaon_mechanical_partonic}
\end{figure*}

The two continuations give noticeably different values
of the central pressure. Approximately
$0.318\,\mathrm{GeV/fm}^2$ of $p_K(0)$ comes from the
directly calculated interval, while the ultraviolet
tails contribute $2.312$ and $1.234\,\mathrm{GeV/fm}^2$
for Ansatz 1 and Ansatz 2, respectively. The resulting
central pressures are $2.6293$ and
$1.5516\,\mathrm{GeV/fm}^2$. Thus, the central pressure
is particularly sensitive to the form-factor behavior
above the calculated momentum-transfer range. The first nonzero-distance pressure zero occurs at
$b_{0,p}=0.2720\,\mathrm{fm}$ for Ansatz 1 and
$0.2906\,\mathrm{fm}$ for Ansatz 2. The corresponding
transverse mechanical radii,
$\sqrt{\langle b_\perp^2\rangle_{\rm mech}}$, are
$0.4616$ and $0.4768\,\mathrm{fm}$, respectively.
For the weighted pressure $2\pi b_\perp p_K$, which
contains no additional factor of $M_K$, the peak values
are $0.5949$ and $0.5395\,\mathrm{GeV/fm}$.
The maximum shear stresses are $1.1929$ and
$0.8440\,\mathrm{GeV/fm}^2$, respectively. The pressure
zero and mechanical radius therefore show much smaller
changes between the two continuations than the central
pressure.

The factor $2\pi b_\perp$ makes the weighted distributions
vanish at the origin; $b_{0,p}$ denotes the first
nonzero-distance zero of the pressure. The transverse
equilibrium relation in Eq.~\eqref{eq:localEquilibrium}
provides a check of the Fourier-Bessel transforms.
A comparison with other stress calculations
\cite{ShanahanDetmold2019Pressure} requires consistent
density and dimensional conventions
\cite{KimKim2021LF,FreeseMiller2021,
PolyakovSchweitzer2018,LiVary2024PionStress}.
In particular, the BLFQ quantity
$2\pi b_\perp P^+p^{\rm LF}$ corresponds in our
convention to $2\pi b_\perp M_Kp_K$.

\begin{figure*}[!tbp]
    \centering

    \begin{subfigure}[t]{0.48\textwidth}
        \centering
        \paperfigure{\linewidth}{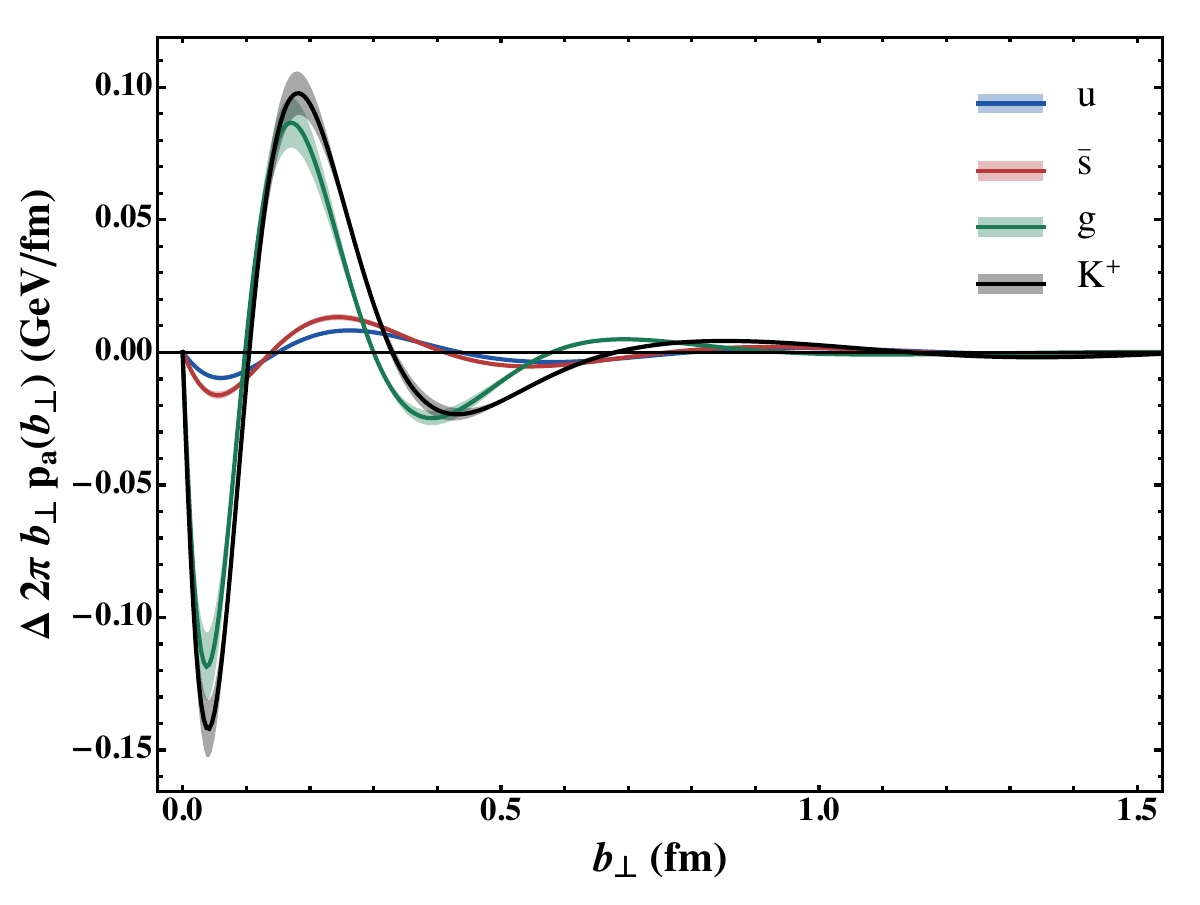}
        \caption{}
        \label{fig:uvdiff_pressure}
    \end{subfigure}
    \hfill
    \begin{subfigure}[t]{0.48\textwidth}
        \centering
        \paperfigure{\linewidth}{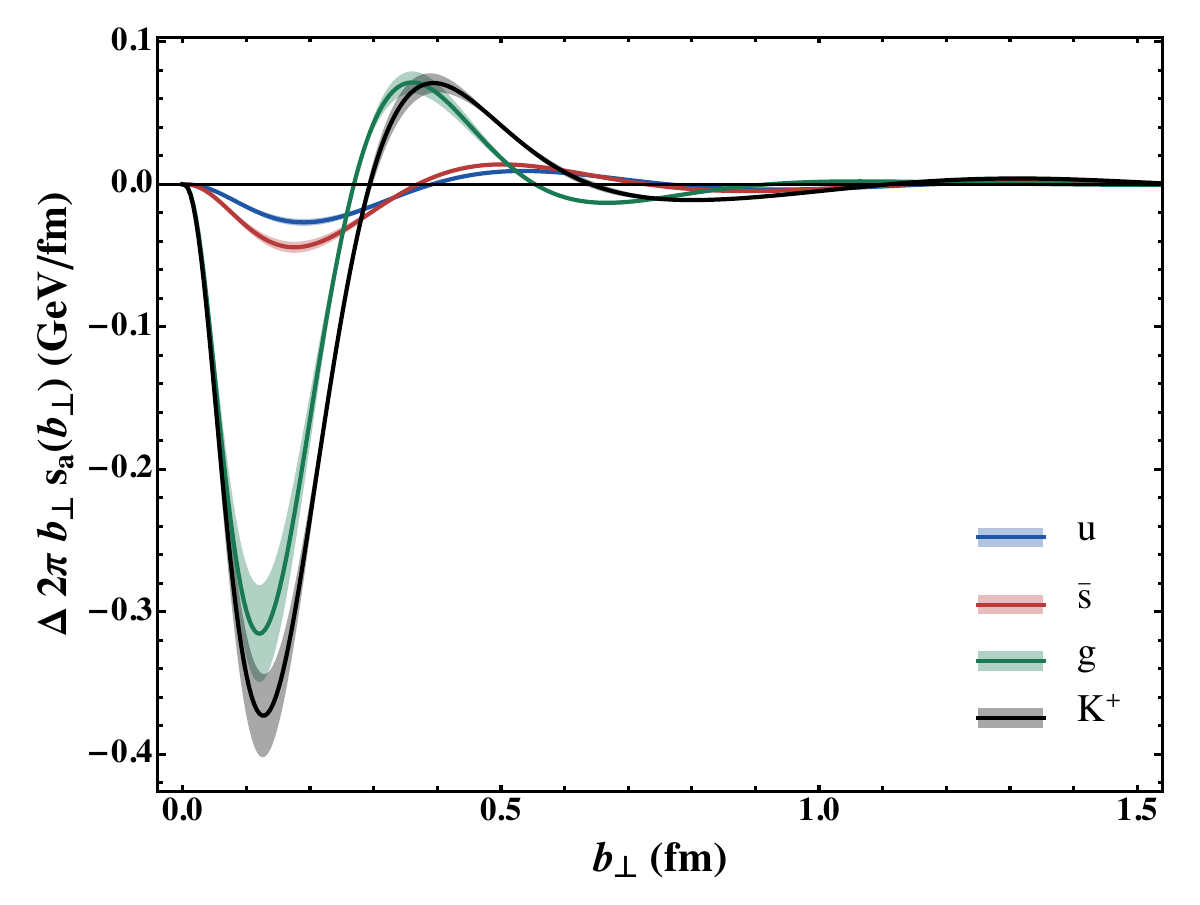}
        \caption{}
        \label{fig:uvdiff_shear}
    \end{subfigure}

    \vspace{0.3em}

    \begin{subfigure}[t]{0.48\textwidth}
        \centering
        \paperfigure{\linewidth}{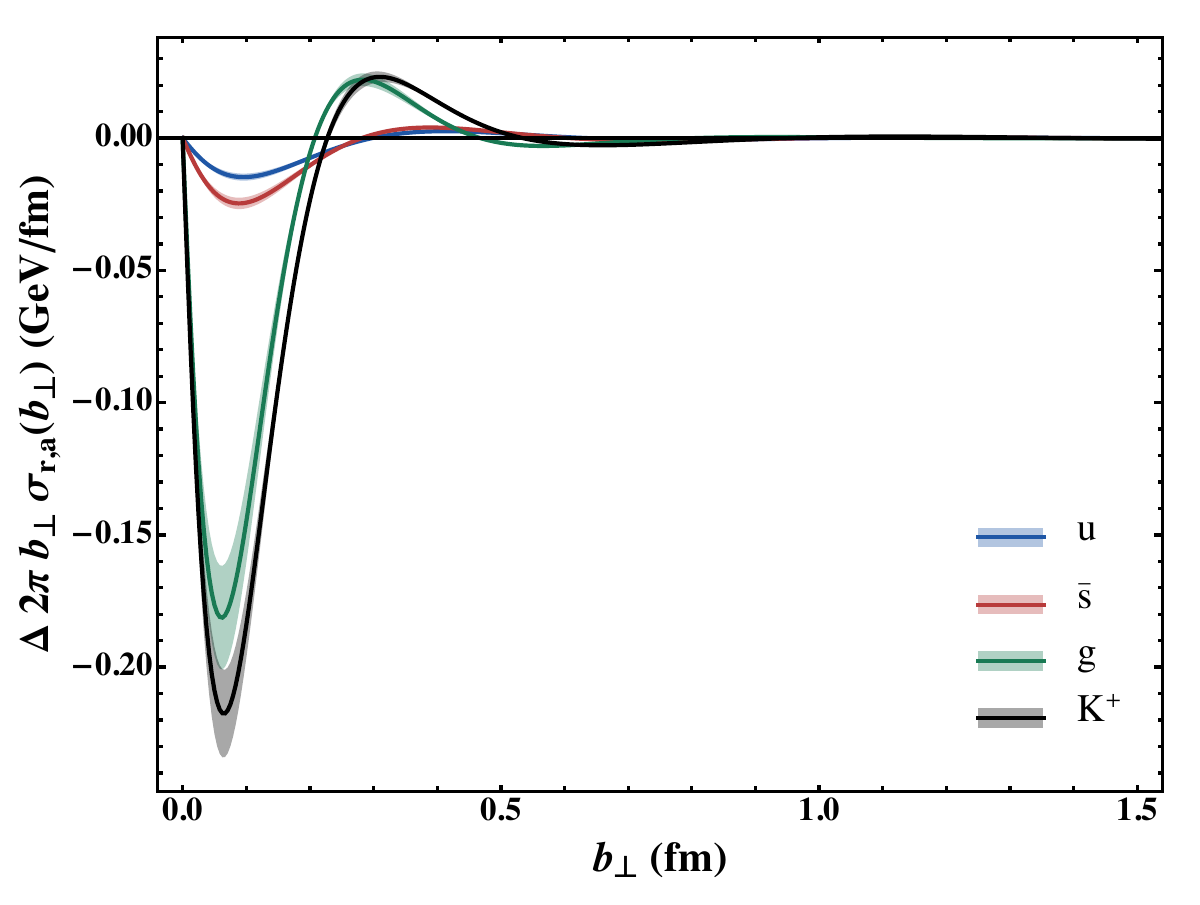}
        \caption{}
        \label{fig:uvdiff_radial}
    \end{subfigure}
    \hfill
    \begin{subfigure}[t]{0.48\textwidth}
        \centering
        \paperfigure{\linewidth}{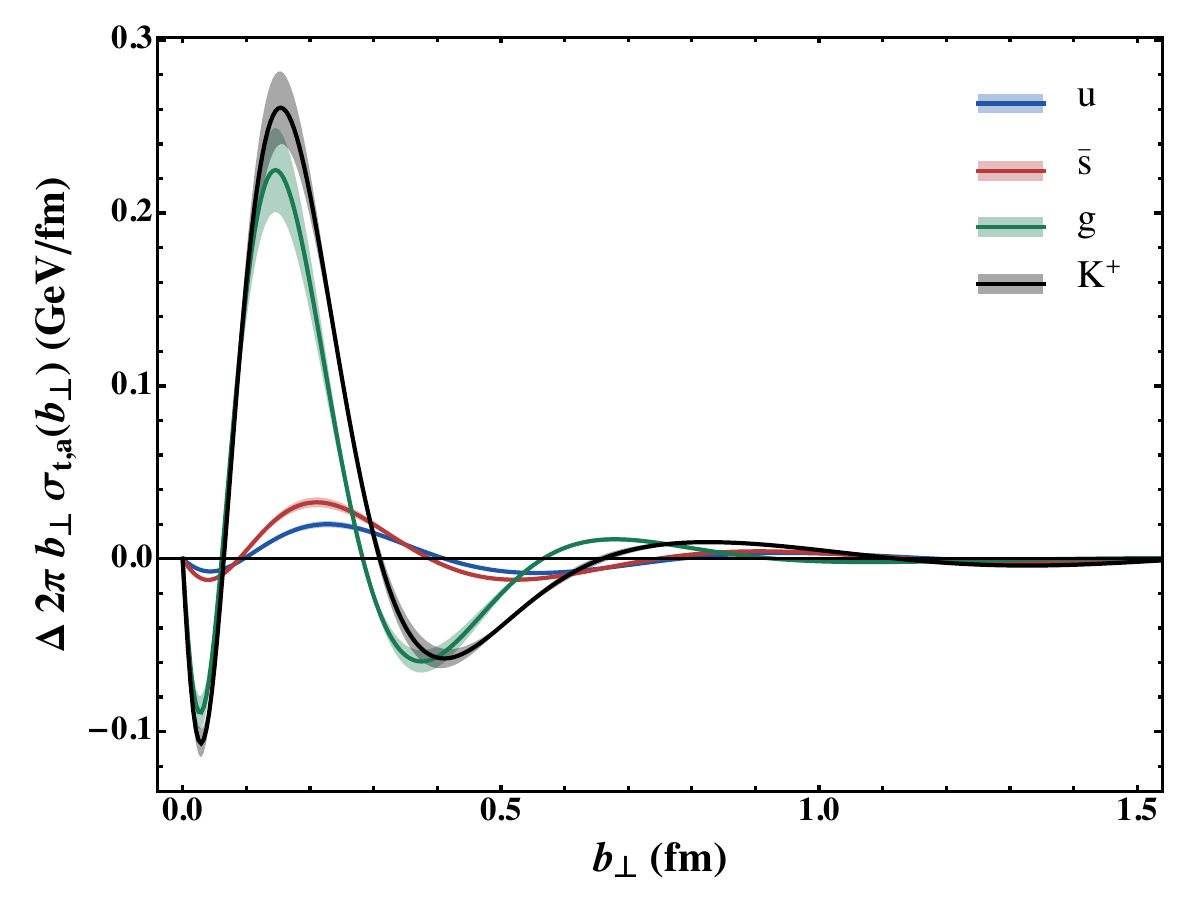}
        \caption{}
        \label{fig:uvdiff_tangential}
    \end{subfigure}

    \caption{Differences between the two ultraviolet
    continuations, $2\pi b_\perp[X_a^{(2)}-X_a^{(1)}]$,
    for the weighted transverse pressure (a), shear (b),
    radial stress (c), and tangential stress (d).
    Colours follow Fig.~\ref{fig:kaon_mechanical_partonic}.
    The bands show the matched mass and basis variations.}
    \label{fig:kaon_mechanical_UV_difference}
\end{figure*}

We next examine the separate partonic contributions.
Figure~\ref{fig:kaon_mechanical_partonic} shows the
weighted pressure, shear, radial-stress, and
tangential-stress distributions for Ansatz 1. The
gluon contribution dominates the peaks of the
pressure, shear, and radial stress, as well as the
negative minimum of the tangential stress. The quark
contributions are smaller, with the strange antiquark
giving the larger positive peaks. The reconstructed
central pressures are approximately
$0.163$, $0.251$, and $2.216\,\mathrm{GeV/fm}^2$
for the $u$, $\bar s$, and gluon contributions,
respectively. The stronger gluon contribution to the
central pressure, compared with its share of $D_K(0)$,
arises from the greater weight of high momentum
transfers in the central-pressure integral. 
The dependence of the mechanical distributions on the
ultraviolet continuation is shown in
Fig.~\ref{fig:kaon_mechanical_UV_difference}, where
we plot the differences between Ansatz 2 and Ansatz 1.
The largest differences occur at
$b_\perp\lesssim0.5\,\mathrm{fm}$ and are mainly
due to the gluon contribution. The differences change
sign at intermediate distances and become small beyond
about $1\,\mathrm{fm}$. Thus, the two continuations
give similar distributions at larger distances but
differ in the short-distance stress profiles.

\subsection{Comparison with other kaon calculations}
\label{sec:resultsComparison}

Table~\ref{tab:kaon_radii_comparison} collects the charged-kaon
radii obtained in this work and in other theoretical and
experimental studies. The comparison includes transverse
and three-dimensional momentum radii, mechanical radii,
pressure-form-factor slope radii, and the electromagnetic
charge radius. Since these quantities are defined through
different form factors and dimensional conventions, their
definitions must be distinguished when comparing the
numerical results. In particular, the normal-force-weighted
mechanical radius is different from the slope-derived
pressure-form-factor radius.

{\renewcommand{\topfraction}{0.95}
\renewcommand{\bottomfraction}{0.95}
\renewcommand{\textfraction}{0.05}
\begin{table}[!t]
\centering
\caption{Charged-kaon radii in fm, grouped by observable
and dimensional convention. Uncertainties for this work
combine mass and basis sensitivities; the two ultraviolet
ansätze are listed separately. Mechanical radii are
normal-force weighted and are distinct from
pressure-form-factor slope radii.}
\label{tab:kaon_radii_comparison}

\footnotesize
\setlength{\tabcolsep}{4pt}
\renewcommand{\arraystretch}{0.95}

\begin{ruledtabular}
\begin{tabular}{@{}llc@{}}
Approach & Observable & Radius (fm)\\
\hline
\multicolumn{3}{l}{\textit{Transverse momentum radius}}\\
This work & $r_{A,\perp}$ & $0.3394\pm0.0113$\\
BLFQ-NJL \cite{Sain2026KaonBLFQ} & $r_{A,\perp}$ & $0.31\pm0.03$\\
\hline
\multicolumn{3}{l}{\textit{Transverse mechanical radius}}\\
This work, Ansatz 1 & $r_{\mathrm{mech},\perp}$ & $0.4616\pm0.0130$\\
This work, Ansatz 2 & $r_{\mathrm{mech},\perp}$ & $0.4768\pm0.0132$\\
BLFQ-NJL \cite{Sain2026KaonBLFQ} & $r_{\mathrm{mech},\perp}$ & $0.55\pm0.06$\\
\hline
\multicolumn{3}{l}{\textit{3D mass-radius conventions}}\\
This work & $r_A^{(3D)}$ & $0.4157\pm0.0139$\\
BLFQ-NJL \cite{Sain2026KaonBLFQ} & $r_A^{(3D)}$ & $0.38\pm0.04$\\
CSM \cite{XuDing2024Kaon} & $r_{\theta_2}^{(3D)}$ & $0.40$\\
Algebraic ansatz \cite{XuDing2024Kaon} & $r_{\theta_2}^{(3D)}$ & $0.41\pm0.03$\\
Holographic QCD \cite{LiuWatanabe2025Kaon} & $r_A^{(3D)}$ & $0.35$\\
\hline
\multicolumn{3}{l}{\textit{3D mechanical-radius convention}}\\
This work, Ansatz 1 & $r_{\mathrm{mech}}^{(3D)}$ & $0.5654\pm0.0159$\\
This work, Ansatz 2 & $r_{\mathrm{mech}}^{(3D)}$ & $0.5840\pm0.0161$\\
BLFQ-NJL \cite{Sain2026KaonBLFQ} & $r_{\mathrm{mech}}^{(3D)}$ & $0.67\pm0.07$\\
\hline
\multicolumn{3}{l}{\textit{3D pressure-form-factor slope radius}}\\
This work & $r_{\theta_1}^{(3D)}$ & $0.9602\pm0.1128$\\
CSM \cite{XuDing2024Kaon} & $r_{\theta_1}^{(3D)}$ & $0.63$\\
Algebraic ansatz \cite{XuDing2024Kaon} & $r_{\theta_1}^{(3D)}$ & $0.68\pm0.04$\\
\hline
\multicolumn{3}{l}{\textit{Electromagnetic charge radius}}\\
This work & $r_{\mathrm{ch}}$ & $0.5178\pm0.0278$\\
PDG (2026) \cite{PDG2026Kaon} & $r_{\mathrm{ch}}$ & $0.560\pm0.031$\\
NA7 (1986) \cite{Amendolia1986Kaon} & $r_{\mathrm{ch}}$ & $0.580\pm0.040$\\
Dally et al. (1980) \cite{Dally1980Kaon} & $r_{\mathrm{ch}}$ & $0.530\pm0.050$\\
Dispersive analysis \cite{Stamen2022Kaon} & $r_{\mathrm{ch}}$ & $0.599\pm0.003$\\
CSM \cite{XuDing2024Kaon} & $r_{\mathrm{ch}}$ & $0.58$\\
Algebraic ansatz \cite{XuDing2024Kaon} & $r_{\mathrm{ch}}$ & $0.51\pm0.03$\\
NJL \cite{Gifari2024Kaon} & $r_{\mathrm{ch}}$ & $0.586$\\
\end{tabular}
\end{ruledtabular}
\end{table}}

The transverse momentum radius obtained from the $A$ form
factor is $r_{A,\perp}=0.3394\pm0.0113\,\mathrm{fm}$,
which is close to the BLFQ-NJL result of
$0.31\pm0.03\,\mathrm{fm}$ \cite{Sain2026KaonBLFQ}.
The sensitivity to the slope-extraction window is
$0.0084\,\mathrm{fm}$, separately from the mass and
basis sensitivity quoted above. In the three-dimensional
slope convention, we obtain
$r_A^{(3D)}=0.4157\pm0.0139\,\mathrm{fm}$.
This value is comparable to the BLFQ-NJL result and
the continuum and algebraic-ansatz results reported
in Ref.~\cite{XuDing2024Kaon}, while exceeding the
holographic QCD result \cite{LiuWatanabe2025Kaon}.

For the mechanical structure, the transverse radii are
$0.4616\pm0.0130\,\mathrm{fm}$ and
$0.4768\pm0.0132\,\mathrm{fm}$ for Ansatz 1 and
Ansatz 2, respectively. Their three-dimensional
counterparts are $0.5654\pm0.0159\,\mathrm{fm}$ and
$0.5840\pm0.0161\,\mathrm{fm}$. Both are smaller than
the corresponding central values reported in the
BLFQ-NJL calculation
\cite{Sain2026KaonBLFQ}. The two ultraviolet
continuations give similar mechanical radii, despite
their larger difference in the central pressure.

The pressure-form-factor slope radius is
$r_{\theta_1}^{(3D)}=0.9602\pm0.1128\,\mathrm{fm}$,
which is larger than the CSM and algebraic-ansatz
values reported in Ref.~\cite{XuDing2024Kaon}.
This radius is obtained from the normalized slope
of the pressure form factor near $Q^2=0$ and should
not be identified with the normal-force-weighted
mechanical radius. The two quantities characterize
different aspects of the transverse mechanical
structure.

For the electromagnetic charge radius, we obtain
$r_{\rm ch}=0.5178\pm0.0278\,\mathrm{fm}$. Its central
value is below the PDG and NA7 values, while being
close to the Dally measurement. The mass and basis
sensitivity band overlaps the quoted ranges of the
PDG, NA7, and Dally results. The dispersive value
\cite{Stamen2022Kaon} is higher, as are the central
values from the CSM and NJL calculations. The
algebraic-ansatz result is closer to our value.
Overall, the calculated charge radius is of the same
order as the experimental and theoretical
determinations, although some differences remain
between the central values.

Other approaches to kaon structure include continuum
calculations \cite{XuDing2024Kaon,Yao2025},
dispersive analyses \cite{CaoGuo2025Dispersive},
NJL models \cite{LiuAbuki2026KaonNJL}, contact
interactions \cite{Sultan2024Pseudoscalar,Xing2023Pion},
and chiral constraints \cite{Donoghue1991Pion}.
These calculations provide a broader context for the
present results, although quantitative comparisons
require consistent current prescriptions and radius
definitions.

Lattice calculations provide further benchmarks for
the partonic decomposition
\cite{Alexandrou2025PionKaonMom,
ShanahanDetmold2019Gluon,Hackett2023PionLattice,
Pefkou2022Hadron}. Such comparisons require matching
the scale and the quark and gluon operator
prescriptions. Measurements of gluon contributions
in the proton \cite{Duran2023Gluon} provide related
motivation for studying gluon structure, but do not
directly constrain the kaon gravitational form factors.

\section{Conclusion}
\label{sec:conclusion}

We have studied the electromagnetic and gravitational form
factors of the charged kaon in a light-front Hamiltonian
model with coupled $|u\bar s\rangle$ and $|u\bar s g\rangle$
Fock sectors. The Hamiltonian parameters were calibrated
to the physical kaon mass, and the resulting eigenstates
were used to calculate the electromagnetic form factor
and the separate quark, antiquark, and gluon contributions
to the gravitational form factors and transverse
mechanical distributions. At the effective model scale
$\mu_0=0.580\,\mathrm{GeV}$, determined
through NNLO evolution and matching to the physical-point
lattice-QCD gluon-to-singlet-quark momentum fraction ratio at
$\mu=2\,\mathrm{GeV}$ in the $\overline{\mathrm{MS}}$
scheme, the strange antiquark carries the largest fraction
of the kaon's longitudinal momentum, whereas the gluon
carries only $2.05\%$. Its contribution to the stress form
factor is much larger: the gluon accounts for $42.4\%$ of
$|D_K(0)|$, with $D_K(0)=-2.03610$. This result shows
that a small gluon momentum fraction does not imply a
similarly small contribution to the transverse stress.
Within the adopted projected Belinfante-current
prescription, the interaction and Fock-sector-changing
terms play an important role in the gluon contribution.
The calculated charge and transverse momentum radii are
$0.5178\pm0.0278\,\mathrm{fm}$ and
$0.3394\pm0.0113\,\mathrm{fm}$, respectively.
The transverse mechanical distributions obtained with
the two ultraviolet continuations show positive pressure
at small transverse distances and negative pressure at
larger distances. The corresponding transverse mechanical
radii are $0.4616\pm0.0130\,\mathrm{fm}$ and
$0.4768\pm0.0132\,\mathrm{fm}$. Both are larger than
the transverse momentum radius. Although the two
continuations give comparable pressure zeros and
mechanical radii, they lead to substantially different
central pressures. This sensitivity indicates that
the short-distance mechanical structure is strongly
affected by the form-factor behavior beyond the
momentum-transfer range calculated directly.

An important extension of this work is to include
explicit sea-quark pairs through the
$|u\bar s q\bar q\rangle$ Fock sector. This will require
the corresponding coupled Hamiltonian and current
matrix elements and will allow us to examine the role
of sea quarks in the momentum sharing and mechanical
structure of the kaon. Further calculations at higher
momentum transfers and the evolution of the separate
quark and gluon gravitational form factors to a common
QCD scale, with quark-gluon operator mixing, will be
important for constraining the ultraviolet continuation
and comparing with lattice-QCD results.

\FloatBarrier
\appendix
\section{Interaction kernels and projected Hamiltonian}
\label{app:kernels}

The two emission routes are specified by the active parent momentum and the splitting fraction. For either active line, a parent of fraction $X$ divides into a same-flavor fermion with fraction $zX$ and a gluon with fraction $(1-z)X$. The relative transverse momentum is $\bm\kappa=(1-z)\bm k_f-z\bm k_g$. In the three-body variables, the quark and antiquark routes have
\begin{equation}
 \begin{aligned}
 X_u&=x_1+x_3,&z_u&=x_1/X_u,\\
 X_{\bar s}&=x_2+x_3,&z_{\bar s}&=x_2/X_{\bar s},\\
 \bm\kappa_{f}&=(1-z_f)\bm k_f-z_f\bm k_3.
 \end{aligned}
 \label{eq:splitmap}
\end{equation}
The normalization constant in the kernels below is $C=16\pi^3$. The canonically normalized helicity kernels are
\begin{equation}
 \begin{aligned}
 \Gamma_f(h,h',\sigma)
 &=\frac{\mathcal C_f\sqrt{2/z_f}\,
 Q_{hh'\sigma}(z_f,\bm\kappa_f;m_f)}
 {\sqrt{C\,X_f x_f x_3}},\\
 \widetilde\Gamma_f&=F_{\lambda_V}(\Delta\mathcal M_f^2)\Gamma_f.
 \end{aligned}
 \label{eq:GammaRoutesMain}
\end{equation}
The six nonzero numerators are listed in Table~\ref{tab:vertexHel}; all other helicity amplitudes vanish. The active splitting measure is $X_f\,dz_f\,d^2\kappa_f$. The finite basis matrix elements are obtained by integrating $\widetilde\Gamma_f$ between the analytic wave functions. No separate vertex fit is carried out.
\begin{table}[!htbp]
\caption{Nonzero light-front helicity numerators of the canonical $f\to fg$ vertex. $\kappa^{R,L}=\kappa_x\pm i\kappa_y$ and $h,h'=\pm\tfrac12$ (indicated by signs).}
\label{tab:vertexHel}
\centering
\begin{tabular}{ccc c}
\toprule
$h$&$h'$&$\sigma$&$Q_{hh'\sigma}(z,\bm\kappa;m)$\\
\midrule
$+$&$+$&$+$&$\kappa^L/(1-z)$\\
$+$&$+$&$-$&$-z\kappa^R/(1-z)$\\
$+$&$-$&$+$&$m(1-z)$\\
$-$&$+$&$-$&$m(1-z)$\\
$-$&$-$&$+$&$z\kappa^L/(1-z)$\\
$-$&$-$&$-$&$-\kappa^R/(1-z)$\\
\bottomrule
\end{tabular}
\end{table}

The three-body pair interactions share a common kinematic representation. Let $X_{rs}=x_r+x_s$ and let the incoming active pair have fractions $x_r'=X_{rs}z$ and $x_s'=X_{rs}(1-z)$, keeping the remaining spectator unchanged. For each pair, the transverse momenta are $\bm k_r'=z\bm K_{rs}+\bm\kappa$ and $\bm k_s'=(1-z)\bm K_{rs}-\bm\kappa$, with $\bm K_{rs}=\bm k_r+\bm k_s$. The exact active-pair Jacobian is $X_{rs}\,dz\,d^2\kappa$. With the finite signed kernels of Eq.~\eqref{eq:R12}, the three nonzero helicity-diagonal interactions in $H_{33}$ are
\begin{align}
 \mathcal K_{12}&=\frac{2}{3C}
 R_{2,\epsilon}(x_1-x_1'),
 \label{eq:K12explicit}\\
 \mathcal K_{13}&=\frac1{C\sqrt{x_3x_3'}}\bigg[
 -3(x_3+x_3')R_{2,\epsilon}(x_1-x_1')+\frac{4}{3}(1-2h_1\sigma)R_{1,\epsilon}(X_{13})-\frac16(1+2h_1\sigma)
 R_{1,\epsilon}(x_1'-x_3)\bigg],
 \label{eq:K13explicit}\\
 \mathcal K_{23}&=\frac1{C\sqrt{x_3x_3'}}\bigg[
 -3(x_3+x_3')R_{2,\epsilon}(x_2-x_2')
 +\frac43(1-2h_2\sigma)R_{1,\epsilon}(X_{23})-\frac{1}{6}(1+2h_2\sigma)
 R_{1,\epsilon}(x_2'-x_3)\bigg].
 \label{eq:K23explicit}
\end{align}
Their color factors are fixed by Eq.~\eqref{eq:paircolors}. These expressions are the direct kernels used in $V_{33}^{\rm can}$, including the mixed constrained fermion-gluon terms in the $13$ and $23$ pairs. They are inserted with the pair measure above, the orbital projections of the three-body HO basis and the appropriate unchanged spectator. The canonical factors have not been modified to mimic the effective gluon mass. Deterministic quadrature evaluates the pair integrals while preserving the adjoint relations.

The projected Hamiltonian acts on the basis coefficients as follows. For the valence spin-zero $U$ channel, the canonical instantaneous exchange survives the angular integral,
\begin{equation}
 \begin{aligned}
 (H_{22}U)(x,k)={}&\mathcal M_2^2U+V_{\rm conf}^{(0)}U
 -\frac{32\pi g^2}{3C}\int_0^1\!dz
 \int_0^\infty\!\kappa\,d\kappa
 \times R_{2,\epsilon}(x-z)U(z,\kappa),\\
 (H_{22}V)(x,k)={}&\mathcal M_2^2V+V_{\rm conf}^{(1)}V.
 \end{aligned}
 \label{eq:H22UV}
\end{equation}
The superscript denotes the azimuthal harmonic of confinement. In the three-body channel the quark-antiquark pair is an octet, and the color factors in Eq.~\eqref{eq:paircolors} replace those in the valence sector. All interactions are projected onto the same two- and three-body basis as the kinetic terms. The full numerical action is
\begin{equation}
 \begin{pmatrix}c_2'\\c_3'\end{pmatrix}=
 \begin{pmatrix}
 H_{22}&g\mathcal V_{23}\\
 g\mathcal V_{32}&D_3(m_g)+g^2V_{33}^{\rm can}
 \end{pmatrix}
 \begin{pmatrix}c_2\\c_3\end{pmatrix}.
 \label{eq:ABmatvec}
\end{equation}
The cubic blocks are constructed with the exact splitting kinematics and the adjoint relation $\mathcal V_{23}=\mathcal V_{32}^\dagger$. No additional phenomenological three-body interaction is added. The three direct pair interactions $V_{33}^{\rm can}$ are evaluated in active-pair coordinates, including the nonlocal longitudinal constrained-field kernels, and the full-plane transverse contractions are implemented in factorized form. The EMT insertions evaluated between these states are described in Sec.~\ref{sec:theoryD}.

\section{Analytic transforms of the ultraviolet continuation}
\label{app:transforms}

In our numerical treatment the Fourier-Bessel transform is performed without a hard high-$Q$ cutoff. The rational tail is decomposed into powers of $(1+\tau/A)^{-m}$ and transformed analytically in terms of modified Bessel functions $K_\nu$. The direct interval is integrated by deterministic quadrature and corrected by subtracting the rational contribution over exactly the same interval. For instance, the basic all-range transform is
\begin{equation}
 \begin{aligned}
 \int_0^\infty\!\frac{d\tau\,J_0(b_\perp\sqrt \tau/\hbar c)}{(1+\tau/A)^m}
 &=\frac{2A}{2^{m-1}\Gamma(m)}
 z^{m-1}K_{m-1}(z), \qquad
 z=\frac{b_\perp\sqrt A}{\hbar c}.
 \end{aligned}
 \label{eq:analyticTransforms}
\end{equation}
Thus the only high-$Q^2$ input is the stated analytic continuation, rather than an additional numerical integration cutoff. For a linear Bessel-transform functional $\mathcal L$, the piecewise construction can be evaluated as
\begin{equation}
 \begin{aligned}
 \mathcal L[D_{a,s}^{(n)}]
 ={}&\mathcal L_{[0,\tau_0]}[D_{a,s}^{\rm PCHIP}]+R_{a,s}^{(n)}\bigl\{
 \mathcal L_{[0,\infty)}[D_{a,s}(0)F_{a,n}]-\mathcal L_{[0,\tau_0]}[D_{a,s}(0)F_{a,n}]\bigr\}.
 \end{aligned}
 \label{eq:analyticTailSubtraction}
\end{equation}
where $R_{a,s}^{(n)}=D_{a,s}(-\tau_0)/[D_{a,s}(0)F_{a,n}(\tau_0)]$. The full-range term is analytic, while the two finite-range pieces are computed with the same deterministic quadrature.

\bibliographystyle{apsrev4-2}
\bibliography{kaon_references}

@article{Teryaev2016GFF,
  author = {Teryaev, Oleg V.},
  title = {Gravitational form factors and nucleon spin structure},
  journal = {Front. Phys.},
  volume = {11},
  pages = {111207},
  year = {2016},
  doi = {10.1007/s11467-016-0573-6}
}

@article{Burkardt2000Impact,
  author = {Burkardt, Matthias},
  title = {Impact parameter dependent parton distributions and
           off-forward parton distributions for {$\zeta\to 0$}},
  journal = {Phys. Rev. D},
  volume = {62},
  pages = {071503},
  year = {2000},
  doi = {10.1103/PhysRevD.62.071503},
  eprint = {hep-ph/0005108},
  archivePrefix = {arXiv},
  note = {Erratum: Phys. Rev. D 66, 119903 (2002)}
}

@article{Burkardt2003Impact,
  author = {Burkardt, Matthias},
  title = {Impact parameter space interpretation for generalized
           parton distributions},
  journal = {Int. J. Mod. Phys. A},
  volume = {18},
  pages = {173--208},
  year = {2003},
  doi = {10.1142/S0217751X03012370},
  eprint = {hep-ph/0207047},
  archivePrefix = {arXiv}
}

@misc{Burkert2021Shear,
  author = {Burkert, Volker D. and Elouadrhiri, Latifa
            and Girod, Francois-Xavier},
  title = {Determination of shear forces inside the proton},
  year = {2021},
  eprint = {2104.02031},
  archivePrefix = {arXiv},
  primaryClass = {nucl-ex},
  url = {https://arxiv.org/abs/2104.02031}
}

@article{FreeseMiller2021Empirical,
  author = {Freese, Adam and Miller, Gerald A.},
  title = {Genuine empirical pressure within the proton},
  journal = {Phys. Rev. D},
  volume = {104},
  pages = {014024},
  year = {2021},
  doi = {10.1103/PhysRevD.104.014024},
  eprint = {2104.03213},
  archivePrefix = {arXiv},
  primaryClass = {hep-ph}
}

@article{Roberts2021Mesons,
  author = {Roberts, Craig D. and Richards, David G.
            and Horn, Tanja and Chang, Lei},
  title = {Insights into the emergence of mass from studies
           of pion and kaon structure},
  journal = {Prog. Part. Nucl. Phys.},
  volume = {120},
  pages = {103883},
  year = {2021},
  doi = {10.1016/j.ppnp.2021.103883},
  eprint = {2102.01765},
  archivePrefix = {arXiv},
  primaryClass = {hep-ph}
}

@article{Zhang2021MesonGPD,
  author = {Zhang, Jin-Li and Raya, Kh{\'e}pani
            and Chang, Lei and Cui, Zhu-Fang
            and Morgado, Jose Manuel and Roberts, Craig D.
            and Rodr{\'i}guez-Quintero, Jos{\'e}},
  title = {Measures of pion and kaon structure from
           generalised parton distributions},
  journal = {Phys. Lett. B},
  volume = {815},
  pages = {136158},
  year = {2021},
  doi = {10.1016/j.physletb.2021.136158},
  eprint = {2101.12286},
  archivePrefix = {arXiv},
  primaryClass = {hep-ph}
}

@article{Raya2022MesonGPD,
  author = {Raya, Kh{\'e}pani and Cui, Zhu-Fang
            and Chang, Lei and Morgado, Jose Manuel
            and Roberts, Craig D.
            and Rodr{\'i}guez-Quintero, Jos{\'e}},
  title = {Revealing pion and kaon structure via
           generalised parton distributions},
  journal = {Chin. Phys. C},
  volume = {46},
  pages = {013105},
  year = {2022},
  doi = {10.1088/1674-1137/ac3071},
  eprint = {2109.11686},
  archivePrefix = {arXiv},
  primaryClass = {hep-ph}
}

@article{SonHutauruk2025GPD,
  author = {Son, Hyeon-Dong and Hutauruk, Parada T. P.},
  title = {Generalized parton distributions of the kaon and
           pion within the nonlocal chiral quark model},
  journal = {Phys. Rev. D},
  volume = {111},
  pages = {054007},
  year = {2025},
  doi = {10.1103/PhysRevD.111.054007},
  eprint = {2411.18130},
  archivePrefix = {arXiv},
  primaryClass = {hep-ph}
}

@misc{Mezrag2014PionGPD,
  author = {Mezrag, C{\'e}dric and Moutarde, Herv{\'e}
            and Rodr{\'i}guez-Quintero, Jos{\'e}
            and Sabati{\'e}, Franck},
  title = {Towards a pion generalized parton distribution
           model from {Dyson--Schwinger} equations},
  year = {2014},
  eprint = {1406.7425},
  archivePrefix = {arXiv},
  primaryClass = {hep-ph},
  url = {https://arxiv.org/abs/1406.7425}
}

@article{Pagels1966EMT,
  author = {Pagels, Heinz},
  title = {Energy-Momentum Structure Form Factors of Particles},
  journal = {Phys. Rev.},
  volume = {144},
  pages = {1250--1260},
  year = {1966},
  doi = {10.1103/PhysRev.144.1250}
}

@article{Polyakov2003Forces,
  author = {Polyakov, Maxim V.},
  title = {Generalized parton distributions and strong forces
           inside nucleons and nuclei},
  journal = {Phys. Lett. B},
  volume = {555},
  pages = {57--62},
  year = {2003},
  doi = {10.1016/S0370-2693(03)00036-4},
  eprint = {hep-ph/0210165},
  archivePrefix = {arXiv},
  primaryClass = {hep-ph}
}

@article{BelitskyRadyushkin2005,
  author = {Belitsky, Andrei V. and Radyushkin, Anatoly V.},
  title = {Unraveling hadron structure with generalized
           parton distributions},
  journal = {Phys. Rept.},
  volume = {418},
  pages = {1--387},
  year = {2005},
  doi = {10.1016/j.physrep.2005.06.002},
  eprint = {hep-ph/0504030},
  archivePrefix = {arXiv},
  primaryClass = {hep-ph}
}

@article{AbidinCarlson2008Axial,
  author = {Abidin, Zainul and Carlson, Carl E.},
  title = {Gravitational form factors in the axial sector
           from an {AdS/QCD} model},
  journal = {Phys. Rev. D},
  volume = {77},
  pages = {115021},
  year = {2008},
  doi = {10.1103/PhysRevD.77.115021},
  eprint = {0804.0214},
  archivePrefix = {arXiv},
  primaryClass = {hep-ph}
}

@article{Diehl2001Overlap,
  author = {Diehl, Markus and Feldmann, Thorsten
            and Jakob, Rainer and Kroll, Peter},
  title = {The overlap representation of skewed quark
           and gluon distributions},
  journal = {Nucl. Phys. B},
  volume = {596},
  pages = {33--65},
  year = {2001},
  eprint = {hep-ph/0009255},
  archivePrefix = {arXiv},
  note = {Erratum: Nucl. Phys. B 605, 647 (2001)}
}

@article{Lan2022DynamicalGluon,
  author = {Lan, Jiangshan and Fu, Kaiyu and Mondal, Chandan
            and Zhao, Xingbo and Vary, James P.},
  title = {Light mesons with one dynamical gluon on the light front},
  journal = {Phys. Lett. B},
  volume = {825},
  pages = {136890},
  year = {2022},
  doi = {10.1016/j.physletb.2022.136890}
}

@article{JiMaYuan2004,
  author = {Ji, Xiangdong and Ma, Jian-Ping and Yuan, Feng},
  title = {Classification and asymptotic scaling of hadrons'
           light-cone wave-function amplitudes},
  journal = {Eur. Phys. J. C},
  volume = {33},
  pages = {75--90},
  year = {2004},
  doi = {10.1140/epjc/s2003-01563-y},
  eprint = {hep-ph/0304107},
  archivePrefix = {arXiv},
  primaryClass = {hep-ph}
}

@article{Haber2021Color,
  author = {Haber, Howard E.},
  title = {Useful relations among the generators in the defining
           and adjoint representations of {SU(N)}},
  journal = {SciPost Phys. Lect. Notes},
  volume = {21},
  year = {2021},
  doi = {10.21468/SciPostPhysLectNotes.21},
  eprint = {1912.13302},
  archivePrefix = {arXiv},
  primaryClass = {math-ph}
}

@article{Glazek2012Effective,
  author = {G{\l}azek, Stanis{\l}aw D.},
  title = {Perturbative formulae for relativistic interactions
           of effective particles},
  journal = {Acta Phys. Polon. B},
  volume = {43},
  pages = {1843--1862},
  year = {2012},
  doi = {10.5506/APhysPolB.43.1843}
}

@article{alexandrou2022scalar,
  title={Scalar, vector, and tensor form factors for the pion and kaon from lattice QCD},
  author={Alexandrou, Constantia and Bacchio, Simone and Cloet, Ian and Constantinou, Martha and Delmar, Joseph and Hadjiyiannakou, Kyriakos and Koutsou, Giannis and Lauer, Colin and Vaquero, Alejandro and (ETM Collaboration)},
  journal={Physical Review D},
  volume={105},
  number={5},
  pages={054502},
  year={2022},
  publisher={APS}
}

@misc{DLMFOrthogonalPolynomials,
  author = {{National Institute of Standards and Technology}},
  title = {{NIST Digital Library of Mathematical Functions}},
  year = {2026},
  howpublished = {\url{https://dlmf.nist.gov/}},
  note = {Release 1.2.8, Sections 18.3 and 18.8.
          Accessed 4 November 2026},
  url = {https://dlmf.nist.gov/}
}

@article{PDG2026Kaon,
  author  = {F. Takahashi and others},
  collaboration = {Particle Data Group},
  title   = {Review of Particle Physics},
  journal = {Int. J. Mod. Phys. A},
  volume  = {41},
  pages   = {2630011},
  year    = {2026},
  note    = {Charged-kaon charge-radius listing}
}

@article{Amendolia1986Kaon,
  author  = {S. R. Amendolia and others},
  title   = {A measurement of the kaon charge radius},
  journal = {Phys. Lett. B},
  volume  = {178},
  pages   = {435--440},
  year    = {1986},
  doi     = {10.1016/0370-2693(86)91407-3}
}

@article{Dally1980Kaon,
  author  = {E. B. Dally and others},
  title   = {Direct Measurement of the Negative-Kaon Form Factor},
  journal = {Phys. Rev. Lett.},
  volume  = {45},
  pages   = {232},
  year    = {1980},
  doi     = {10.1103/PhysRevLett.45.232}
}

@article{Stamen2022Kaon,
  author  = {Dominik Stamen and Deepti Hariharan
             and Martin Hoferichter and Bastian Kubis
             and Peter Stoffer},
  title   = {Kaon electromagnetic form factors in dispersion theory},
  journal = {Eur. Phys. J. C},
  volume  = {82},
  pages   = {432},
  year    = {2022},
  eprint  = {2202.11106},
  archivePrefix = {arXiv},
  doi     = {10.1140/epjc/s10052-022-10348-3}
}

@article{Gifari2024Kaon,
  author  = {Geoffry Gifari and Parada T. P. Hutauruk
             and Terry Mart},
  title   = {Nuclear medium meson structures from the
             Schwinger proper-time Nambu--Jona-Lasinio model},
  journal = {Phys. Rev. D},
  volume  = {110},
  pages   = {014043},
  year    = {2024},
  doi     = {10.1103/PhysRevD.110.014043}
}

@misc{Sain2026KaonBLFQ,
  title = {Gravitational form factors of light mesons from Basis Light-Front Quantization},
  author = {Amrita Sain and Sreeraj Nair and Chandan Mondal and Xingbo Zhao and James P. Vary},
  year = {2026},
  eprint = {2605.18350},
  archivePrefix = {arXiv},
  url = {https://arxiv.org/abs/2605.18350},
}

@misc{LiuAbuki2026KaonNJL,
  title = {Kaon gravitational form factors and mechanical structure in a three-flavor NJL model},
  author = {Zhibo Liu and Hiroaki Abuki},
  year = {2026},
  eprint = {2608.29848},
  archivePrefix = {arXiv},
  url = {https://arxiv.org/abs/2608.29848},
}

@article{XuDing2024Kaon,
  title = {Pion and kaon electromagnetic and gravitational form factors},
  author = {Yin-Zhen Xu and Minghui Ding and Kh'epani Raya and Craig D. Roberts and Jos'e Rodr'iguez-Quintero and Sebastian M. Schmidt},
  journal = {Eur. Phys. J. C},
  volume = {84},
  pages = {191},
  year = {2024},
  eprint = {2311.14832},
  archivePrefix = {arXiv},
  doi = {10.1140/epjc/s10052-024-12518-x},
  url = {https://arxiv.org/abs/2311.14832},
}

@article{CaoGuo2025Dispersive,
  title = {Gravitational form factors of pions, kaons and nucleons from dispersion relations},
  author = {Xiong-Hui Cao and Feng-Kun Guo and Qu-Zhi Li and Bo-Wen Wu and De-Liang Yao},
  journal = {Eur. Phys. J. Spec. Top.},
  volume = {235},
  pages = {2121--2154},
  year = {2025},
  eprint = {2507.05375},
  archivePrefix = {arXiv},
  doi = {10.1140/epjs/s11734-025-02025-9},
  url = {https://arxiv.org/abs/2507.05375},
}

@article{LiuWatanabe2025Kaon,
  title = {Gravitational form factor of the kaon in holographic QCD},
  author = {Zhibo Liu and Akira Watanabe},
  journal = {Phys. Rev. D},
  volume = {112},
  pages = {114032},
  year = {2025},
  eprint = {2503.18747},
  archivePrefix = {arXiv},
  doi = {10.1103/1wbr-j3r4},
  url = {https://arxiv.org/abs/2503.18747},
}

@article{Hackett2023PionLattice,
  title = {Gravitational form factors of the pion from lattice QCD},
  author = {Daniel C. Hackett and Patrick R. Oare and Dimitra A. Pefkou and Phiala E. Shanahan},
  journal = {Phys. Rev. D},
  volume = {108},
  pages = {114504},
  year = {2023},
  eprint = {2307.11707},
  archivePrefix = {arXiv},
  doi = {10.1103/PhysRevD.108.114504},
  url = {https://arxiv.org/abs/2307.11707},
}

@article{LiVary2024PionStress,
  title = {Stress inside the pion in holographic light-front QCD},
  author = {Yang Li and James P. Vary},
  journal = {Phys. Rev. D},
  volume = {109},
  pages = {L051501},
  year = {2024},
  eprint = {2312.02543},
  archivePrefix = {arXiv},
  doi = {10.1103/PhysRevD.109.L051501},
  url = {https://arxiv.org/abs/2312.02543},
}

@article{LiuShuryak2024I,
  title = {Pion gravitational form factors in the QCD instanton vacuum. I},
  author = {Wei-Yang Liu and Edward Shuryak and Christian Weiss and Ismail Zahed},
  journal = {Phys. Rev. D},
  volume = {110},
  pages = {054021},
  year = {2024},
  eprint = {2405.14026},
  archivePrefix = {arXiv},
  doi = {10.1103/PhysRevD.110.054021},
  url = {https://arxiv.org/abs/2405.14026},
}

@article{LiuShuryak2024II,
  title = {Pion gravitational form factors in the QCD instanton vacuum. II},
  author = {Wei-Yang Liu and Edward Shuryak and Ismail Zahed},
  journal = {Phys. Rev. D},
  volume = {110},
  pages = {054022},
  year = {2024},
  eprint = {2405.16269},
  archivePrefix = {arXiv},
  doi = {10.1103/PhysRevD.110.054022},
  url = {https://arxiv.org/abs/2405.16269},
}

@article{Broniowski2024MesonDominance,
  title = {Gravitational form factors of the pion and meson dominance},
  author = {Wojciech Broniowski and Enrique Ruiz Arriola},
  journal = {Phys. Lett. B},
  volume = {859},
  pages = {139138},
  year = {2024},
  eprint = {2405.07815},
  archivePrefix = {arXiv},
  doi = {10.1016/j.physletb.2024.139138},
  url = {https://arxiv.org/abs/2405.07815},
}

@article{Xing2023Pion,
  title = {A glimpse into pion gravitational form factor},
  author = {Zanbin Xing and Minghui Ding and Lei Chang},
  journal = {Phys. Rev. D},
  volume = {107},
  pages = {L031502},
  year = {2023},
  eprint = {2211.06635},
  archivePrefix = {arXiv},
  doi = {10.1103/PhysRevD.107.L031502},
  url = {https://arxiv.org/abs/2211.06635},
}

@article{Sultan2024Pseudoscalar,
  title = {Gravitational form factors of pseudoscalar mesons in a contact interaction},
  author = {M. Atif Sultan and Zanbin Xing and Kh'epani Raya and Adnan Bashir and Lei Chang},
  journal = {Phys. Rev. D},
  volume = {110},
  pages = {054034},
  year = {2024},
  eprint = {2407.10437},
  archivePrefix = {arXiv},
  doi = {10.1103/PhysRevD.110.054034},
  url = {https://arxiv.org/abs/2407.10437},
}

@article{Choi2025PionLFQM,
  title = {Gravitational form factors of the pion in the self-consistent light-front quark model},
  author = {Yongwoo Choi and Hyeon-Dong Son and Ho-Meoyng Choi},
  journal = {Phys. Rev. D},
  volume = {112},
  pages = {014043},
  year = {2025},
  eprint = {2504.14997},
  archivePrefix = {arXiv},
  url = {https://arxiv.org/abs/2504.14997},
}

@article{Donoghue1991Pion,
  title = {Energy and momentum in chiral theories},
  author = {John F. Donoghue and Heinrich Leutwyler},
  journal = {Z. Phys. C},
  volume = {52},
  pages = {343},
  year = {1991},
}

@article{Broniowski2008Pion,
  title = {Gravitational and higher-order form factors of the pion in chiral quark models},
  author = {Wojciech Broniowski and Enrique Ruiz Arriola},
  journal = {Phys. Rev. D},
  volume = {78},
  pages = {094011},
  year = {2008},
  doi = {10.1103/PhysRevD.78.094011},
  url = {https://doi.org/10.1103/PhysRevD.78.094011},
}

@article{CaoLiVary2023Scalar,
  title = {Forces inside a strongly-coupled scalar nucleon},
  author = {Xiong-Hui Cao and Yang Li and James P. Vary},
  journal = {Phys. Rev. D},
  volume = {108},
  pages = {056026},
  year = {2023},
  eprint = {2308.06812},
  archivePrefix = {arXiv},
  doi = {10.1103/PhysRevD.108.056026},
  url = {https://arxiv.org/abs/2308.06812},
}

@article{Sain2025Gluon,
  title = {Gluon gravitational form factors of the proton in a light-front spectator model},
  author = {Amrita Sain and Poonam Choudhary and Bheemsehan Gurjar and Chandan Mondal and Dipankar Chakrabarti and Asmita Mukherjee},
  journal = {Phys. Rev. D},
  volume = {111},
  pages = {094011},
  year = {2025},
  eprint = {2503.12574},
  archivePrefix = {arXiv},
  doi = {10.1103/PhysRevD.111.094011},
  url = {https://arxiv.org/abs/2503.12574},
}

@misc{More2023GluonDressed,
  title = {Gluon contribution to the mechanical properties of a dressed quark in light-front Hamiltonian QCD},
  author = {Jai More and Asmita Mukherjee and Sreeraj Nair and Sreedevi Saha},
  year = {2023},
  eprint = {2302.11906},
  archivePrefix = {arXiv},
  url = {https://arxiv.org/abs/2302.11906},
}

@article{PolyakovSchweitzer2018,
  title = {Forces inside hadrons: pressure, surface tension, mechanical radius, and all that},
  author = {Maxim V. Polyakov and Peter Schweitzer},
  journal = {Int. J. Mod. Phys. A},
  volume = {33},
  pages = {1830025},
  year = {2018},
  eprint = {1805.06596},
  archivePrefix = {arXiv},
  doi = {10.1142/S0217751X18300259},
  url = {https://arxiv.org/abs/1805.06596},
}

@article{BurkertEtAl2023Review,
  author = {Volker D. Burkert and Latifa Elouadrhiri and Francois-Xavier Girod and C'edric Lorc'e and Peter Schweitzer and Phiala E. Shanahan},
  journal = {Rev. Mod. Phys.},
  volume = {95},
  pages = {041002},
  year = {2023},
  eprint = {2303.08347},
  archivePrefix = {arXiv},
  url = {https://arxiv.org/abs/2303.08347},
}

@article{FreeseMiller2021,
  title = {Forces within hadrons on the light front},
  author = {Adam Freese and Gerald A. Miller},
  journal = {Phys. Rev. D},
  volume = {103},
  pages = {094023},
  year = {2021},
  eprint = {2102.01683},
  archivePrefix = {arXiv},
  doi = {10.1103/PhysRevD.103.094023},
  url = {https://arxiv.org/abs/2102.01683},
}

@article{Lorce2019Mechanics,
  title = {Revisiting the mechanical properties of the nucleon},
  author = {C'edric Lorc'e and Herv'e Moutarde and Artur P. Trawi'nski},
  journal = {Eur. Phys. J. C},
  volume = {79},
  pages = {89},
  year = {2019},
  eprint = {1810.09837},
  archivePrefix = {arXiv},
  doi = {10.1140/epjc/s10052-019-6572-3},
  url = {https://arxiv.org/abs/1810.09837},
}

@article{KimKim2021LF,
  author = {Jung-Yun Kim and Hyun-Chul Kim},
  journal = {Phys. Rev. D},
  volume = {104},
  pages = {074019},
  year = {2021},
  eprint = {2105.10279},
  archivePrefix = {arXiv},
  doi = {10.1103/PhysRevD.104.074019},
  url = {https://arxiv.org/abs/2105.10279},
}

@article{Ji1997GPD,
  title = {Deeply virtual Compton scattering},
  author = {Xiangdong Ji},
  journal = {Phys. Rev. D},
  volume = {55},
  pages = {7114},
  year = {1997},
  eprint = {hep-ph/9609381},
  archivePrefix = {arXiv},
  doi = {10.1103/PhysRevD.55.7114},
  url = {https://arxiv.org/abs/hep-ph/9609381},
}

@article{Radyushkin1997DVCS,
  author = {Anatoly V. Radyushkin},
  journal = {Phys. Rev. D},
  volume = {56},
  pages = {5524},
  year = {1997},
  eprint = {hep-ph/9704207},
  archivePrefix = {arXiv},
  doi = {10.1103/PhysRevD.56.5524},
  url = {https://arxiv.org/abs/hep-ph/9704207},
}

@article{Diehl2003Review,
  title = {Generalized parton distributions},
  author = {Markus Diehl},
  journal = {Phys. Rept.},
  volume = {388},
  pages = {41},
  year = {2003},
  eprint = {hep-ph/0307382},
  archivePrefix = {arXiv},
  doi = {10.1016/j.physrep.2003.08.002},
  url = {https://arxiv.org/abs/hep-ph/0307382},
}

@article{PolyakovSon2018,
  title={Nucleon gravitational form factors from instantons: forces between quark and gluon subsystems},
  author={Polyakov, Maxim V and Son, Hyeon-Dong},
  journal={Journal of High Energy Physics},
  volume={2018},
  number={9},
  pages={156},
  year={2018},
  publisher={Springer},
  doi={10.1007/JHEP09(2018)156}
}

@article{HattaRajanTanaka2018,
  title={Quark and gluon contributions to the QCD trace anomaly},
  author={Hatta, Yoshitaka and Rajan, Abha and Tanaka, Kazuhiro},
  journal={Journal of High Energy Physics},
  volume={2018},
  number={12},
  pages={8},
  year={2018},
  publisher={Springer}
}

@article{BrodskyDeTeramond2008GFF,
  title = {Light-Front Dynamics and AdS/QCD Correspondence: Gravitational Form Factors of Composite Hadrons},
  author = {Stanley J. Brodsky and Guy F. de T'eramond},
  journal = {Phys. Rev. D},
  volume = {78},
  pages = {025032},
  year = {2008},
  eprint = {0804.0452},
  archivePrefix = {arXiv},
  doi = {10.1103/PhysRevD.78.025032},
  url = {https://arxiv.org/abs/0804.0452},
}

@article{BrodskyPauliPinsky1998,
  title = {Quantum chromodynamics and other field theories on the light cone},
  author = {Stanley J. Brodsky and Hans-Christian Pauli and Stephen S. Pinsky},
  journal = {Phys. Rept.},
  volume = {301},
  pages = {299--486},
  year = {1998},
  eprint = {hep-ph/9705477},
  archivePrefix = {arXiv},
  doi = {10.1016/S0370-1573(97)00089-6},
  url = {https://arxiv.org/abs/hep-ph/9705477},
}

@article{Vary2010BLFQ,
  title = {Hamiltonian light-front field theory in a basis function approach},
  author = {James P. Vary and others},
  journal = {Phys. Rev. C},
  volume = {81},
  pages = {035205},
  year = {2010},
  eprint = {0905.1411},
  archivePrefix = {arXiv},
  doi = {10.1103/PhysRevC.81.035205},
  url = {https://arxiv.org/abs/0905.1411},
}

@article{Brodsky2015Holography,
  title = {Light-front holographic QCD and emerging confinement},
  author = {Stanley J. Brodsky and Guy F. de T'eramond and Hans G. Dosch and Joshua Erlich},
  journal = {Phys. Rept.},
  volume = {584},
  pages = {1--105},
  year = {2015},
  eprint = {1407.8131},
  archivePrefix = {arXiv},
  doi = {10.1016/j.physrep.2015.05.001},
  url = {https://arxiv.org/abs/1407.8131},
}

@article{LiMarisZhao2016,
  title = {Heavy quarkonium in a holographic basis},
  author = {Yang Li and Pieter Maris and Xingbo Zhao and James P. Vary},
  journal = {Phys. Lett. B},
  volume = {758},
  pages = {118--124},
  year = {2016},
  eprint = {1509.07212},
  archivePrefix = {arXiv},
  doi = {10.1016/j.physletb.2016.04.065},
  url = {https://arxiv.org/abs/1509.07212},
}

@article{LiMarisVary2017,
  author = {Yang Li and Pieter Maris and James P. Vary},
  journal = {Phys. Rev. D},
  volume = {96},
  pages = {016022},
  year = {2017},
  eprint = {1704.06968},
  archivePrefix = {arXiv},
  doi = {10.1103/PhysRevD.96.016022},
  url = {https://arxiv.org/abs/1704.06968},
}

@article{JiaVary2019NJL,
  author = {Shaoyang Jia and James P. Vary},
  journal = {Phys. Rev. C},
  volume = {99},
  pages = {035206},
  year = {2019},
  eprint = {1811.08512},
  archivePrefix = {arXiv},
  doi = {10.1103/PhysRevC.99.035206},
  url = {https://arxiv.org/abs/1811.08512},
}

@article{Lan2020Kaon,
  author = {Jiangshan Lan and Chandan Mondal and Shaoyang Jia and Xingbo Zhao and James P. Vary},
  journal = {Phys. Rev. D},
  volume = {101},
  pages = {034024},
  year = {2020},
  eprint = {1907.01509},
  archivePrefix = {arXiv},
  doi = {10.1103/PhysRevD.101.034024},
  url = {https://arxiv.org/abs/1907.01509},
}

@article{BrodskyHwang2001,
  title = {Light cone representation of the spin and orbital angular momentum of relativistic composite systems},
  author = {Stanley J. Brodsky and Dae Sung Hwang and Bo-Qiang Ma and Ivan Schmidt},
  journal = {Nucl. Phys. B},
  volume = {593},
  pages = {311--335},
  year = {2001},
  eprint = {hep-th/0003082},
  archivePrefix = {arXiv},
  doi = {10.1016/S0550-3213(00)00626-X},
  url = {https://arxiv.org/abs/hep-th/0003082},
}

@article{Miller2007Transverse,
  title = {Charge density of the neutron},
  author = {Gerald A. Miller},
  journal = {Phys. Rev. Lett.},
  volume = {99},
  pages = {112001},
  year = {2007},
  eprint = {0705.2409},
  archivePrefix = {arXiv},
  doi = {10.1103/PhysRevLett.99.112001},
  url = {https://arxiv.org/abs/0705.2409},
}

@article{Alexandrou2025PionKaonMom,
  title = {Quark and Gluon Momentum Fractions in the Pion and in the Kaon},
  author = {Constantia Alexandrou and others},
  journal = {Phys. Rev. Lett.},
  volume = {134},
  pages = {131902},
  year = {2025},
  doi = {10.1103/PhysRevLett.134.131902},
  url = {https://doi.org/10.1103/PhysRevLett.134.131902},
}

@article{ShanahanDetmold2019Gluon,
  title = {Gluon gravitational form factors of the nucleon and the pion from lattice QCD},
  author = {Phiala E. Shanahan and William Detmold},
  journal = {Phys. Rev. D},
  volume = {99},
  pages = {014511},
  year = {2019},
  eprint = {1810.04626},
  archivePrefix = {arXiv},
  doi = {10.1103/PhysRevD.99.014511},
  url = {https://arxiv.org/abs/1810.04626},
}

@article{ShanahanDetmold2019Pressure,
  title = {Pressure Distribution and Shear Forces inside the Proton},
  author = {Phiala E. Shanahan and William Detmold},
  journal = {Phys. Rev. Lett.},
  volume = {122},
  pages = {072003},
  year = {2019},
  eprint = {1810.07589},
  archivePrefix = {arXiv},
  doi = {10.1103/PhysRevLett.122.072003},
  url = {https://arxiv.org/abs/1810.07589},
}

@article{Pefkou2022Hadron,
  title = {Gluon gravitational structure of hadrons of different spin},
  author = {Dimitra A. Pefkou and Daniel C. Hackett and Phiala E. Shanahan},
  journal = {Phys. Rev. D},
  volume = {105},
  pages = {054509},
  year = {2022},
  eprint = {2107.10368},
  archivePrefix = {arXiv},
  doi = {10.1103/PhysRevD.105.054509},
  url = {https://arxiv.org/abs/2107.10368},
}

@article{Yang2018PionGluon,
  title = {Nonperturbatively renormalized glue momentum fraction at the physical pion mass from lattice QCD},
  author = {Yi-Bo Yang and others},
  journal = {Phys. Rev. D},
  volume = {98},
  pages = {074506},
  year = {2018},
  eprint = {1805.00531},
  archivePrefix = {arXiv},
  doi = {10.1103/PhysRevD.98.074506},
  url = {https://arxiv.org/abs/1805.00531},
}

@article{Alexandrou2020Momentum,
  title = {Complete flavor decomposition of the spin and momentum fraction of the proton using lattice QCD simulations at physical pion mass},
  author = {Constantia Alexandrou and others},
  journal = {Phys. Rev. D},
  volume = {101},
  pages = {094513},
  year = {2020},
  eprint = {2003.08486},
  archivePrefix = {arXiv},
  doi = {10.1103/PhysRevD.101.094513},
  url = {https://arxiv.org/abs/2003.08486},
}

@article{SkullerudWilliams2001,
  title = {Quark propagator in the Landau gauge},
  author = {Jon Ivar Skullerud and Anthony G. Williams},
  journal = {Phys. Rev. D},
  volume = {63},
  pages = {054508},
  year = {2001},
  doi = {10.1103/PhysRevD.63.054508},
  url = {https://doi.org/10.1103/PhysRevD.63.054508},
}

@article{Burkert2018Proton,
  title = {The pressure distribution inside the proton},
  author = {Volker D. Burkert and Latifa Elouadrhiri and Francois-Xavier Girod},
  journal = {Nature},
  volume = {557},
  pages = {396--399},
  year = {2018},
  doi = {10.1038/s41586-018-0060-z},
  url = {https://doi.org/10.1038/s41586-018-0060-z},
}

@article{Duran2023Gluon,
  author = {B. Duran and others},
  journal = {Nature},
  volume = {615},
  pages = {813},
  year = {2023},
  eprint = {2207.05212},
  archivePrefix = {arXiv},
  url = {https://arxiv.org/abs/2207.05212},
}

@article{Kumano2018TwoPhoton,
  title = {Hadron tomography by generalized distribution amplitudes in pion-pair production process},
  author = {Susumu Kumano and Qin-Tao Song and Oleg V. Teryaev},
  journal = {Phys. Rev. D},
  volume = {97},
  pages = {014020},
  year = {2018},
  eprint = {1711.08088},
  archivePrefix = {arXiv},
  doi = {10.1103/PhysRevD.97.014020},
  url = {https://arxiv.org/abs/1711.08088},
}

@article{Masuda2016Belle,
  author = {M. Masuda and others},
  journal = {Phys. Rev. D},
  volume = {93},
  pages = {032003},
  year = {2016},
  eprint = {1508.06757},
  archivePrefix = {arXiv},
  doi = {10.1103/PhysRevD.93.032003},
  url = {https://arxiv.org/abs/1508.06757},
}

@article{KawanaiSasaki2015,
  title = {Potential description of charmonium and charmed-strange mesons from lattice QCD},
  author = {Taichi Kawanai and Shoichi Sasaki},
  journal = {Phys. Rev. D},
  volume = {92},
  pages = {094503},
  year = {2015},
  doi = {10.1103/PhysRevD.92.094503},
  url = {https://doi.org/10.1103/PhysRevD.92.094503},
}

@article{Nair2024ProtonBLFQ,
  title = {Gravitational form factors and mechanical properties of quarks in protons: A basis light-front quantization approach},
  author = {Sreeraj Nair and Chandan Mondal and Siqi Xu and Xingbo Zhao and Asmita Mukherjee and James P. Vary},
  journal = {Phys. Rev. D},
  volume = {110},
  pages = {056027},
  year = {2024},
  eprint = {2403.11702},
  archivePrefix = {arXiv},
  doi = {10.1103/PhysRevD.110.056027},
  url = {https://arxiv.org/abs/2403.11702},
}

@article{CaoEtAl2024LFEMT,
  title = {Energy momentum tensor on and off the light cone: exposition with scalar Yukawa theory},
  author = {Xianghui Cao and Siqi Xu and Yang Li and Guangyao Chen and Xingbo Zhao and Vladimir A. Karmanov and James P. Vary},
  journal = {JHEP},
  volume = {2024},
  number = {07},
  pages = {095},
  year = {2024},
  eprint = {2405.06896},
  archivePrefix = {arXiv},
  doi = {10.1007/JHEP07(2024)095},
  url = {https://arxiv.org/abs/2405.06896},
}

@misc{LiWangVary2024EMT,
  title = {Hadronic energy-momentum tensor revisited},
  author = {Yang Li and Qun Wang and James P. Vary},
  year = {2024},
  eprint = {2405.06892},
  archivePrefix = {arXiv},
  url = {https://arxiv.org/abs/2405.06892},
}

@article{Hu2025ShearStress,
  title = {Gravitational form factor D of charmonium from shear stress},
  author = {Tianyang Hu and Xianghui Cao and Siqi Xu and Yang Li and Xingbo Zhao and James P. Vary},
  journal = {Phys. Rev. D},
  volume = {111},
  pages = {074031},
  year = {2025},
  eprint = {2408.09689},
  archivePrefix = {arXiv},
  doi = {10.1103/PhysRevD.111.074031},
  url = {https://arxiv.org/abs/2408.09689},
}

@article{Lan2025PionGluonFock,
  title = {Gluonic contributions to the pion parton distribution functions},
  author = {Jiangshan Lan and Chandan Mondal and Xingbo Zhao and Tobias Frederico and James P. Vary},
  journal = {Phys. Rev. D},
  volume = {111},
  pages = {L111903},
  year = {2025},
}

@article{Yao2025,
  title = {Pion, Kaon and nucleon gravitational form factors},
  author = {Zhao-Qian Yao and Yin-Zhen Xu and Daniele Binosi and Minghui Ding and Zhu-Fang Cui and Kh'epani Raya and Craig D. Roberts and Jos'e Rodr'iguez-Quintero},
  journal = {PoS},
  volume = {QCHSC24},
  pages = {081},
  year = {2025},
  eprint = {2503.23166},
  archivePrefix = {arXiv},
  url = {https://arxiv.org/abs/2503.23166},
}

@article{SrivastavaBrodsky,
  author = {Prem P. Srivastava and Stanley J. Brodsky},
  title = {Light-front-quantized QCD in light-cone gauge},
  journal = {Phys. Rev. D},
  volume = {64},
  pages = {045006},
  year = {2001},
  doi = {10.1103/PhysRevD.64.045006}
}

\end{document}